\documentclass{aa}
\usepackage[colorlinks=true,citecolor=blue,anchorcolor=blue,filecolor=blue,linkcolor=blue]{hyperref}
\usepackage{comment}
\usepackage{graphicx}
\usepackage{amssymb,amsmath}
\usepackage{latexsym}
\usepackage{mathtools}
\usepackage{epsfig}
\usepackage{natbib}
\usepackage{dsfont}
\usepackage{xcolor}
\usepackage{txfonts}
\usepackage{mathrsfs}
\usepackage{multirow}
\usepackage{arydshln}
\DeclareMathOperator{\arctanh}{arctanh}
\begin{document}

   \title{An efficient integrator for stellar dynamics in effective gravity fields based on the isochrone potential}
%
   \author{Alexandre Bougakov
          \and
          Melaine Saillenfest
          \and
          Marc Fouchard
          }
   \authorrunning{Bougakov et al.}
   \institute{LTE, Observatoire de Paris, Université PSL, Sorbonne Université, Université de Lille, LNE, CNRS, 61 Avenue de l’Observatoire, 75014 Paris, France\\
             \email{alexandre.bougakov@obspm.fr}}
   \date{Received 2025-01-24 / Accepted 2025-03-19 }


  \abstract
  {Integrating the motion of stars immersed in some smoothed potential is necessary in many stellar and galactic studies. Previous works have generally used numerical integrators that alternate between linear drifts and velocity kicks (such as the standard Leapfrog scheme). The low efficiency of this approach contrasts with the sophisticated methods developed in other fields such as planetary dynamics, for which integrators alternate between Keplerian drifts and velocity kicks.}
  {Inspired by the splitting methods used in planetary dynamics, we aim to build an efficient integration scheme dedicated to stellar and galactic dynamics.}
  {We took advantage of the peculiar properties of Hénon's isochrone potential to design a new symplectic splitting scheme that can be used to integrate the motion of stars in any gravitational potential. This scheme alternates between isochrone drifts and velocity kicks. As a first application, we considered the motion of a star in a Plummer potential --- an essential constituent of galactic potentials --- and determined the set of integration parameters that provide the best integration efficiency (i.e. the best conservation of total energy for the lowest computational cost).}
  {We derived the analytical solution for all possible kinds of orbits in Hénon's isochrone potential (bound and unbound trajectories) as needed in our integration scheme. Our numerical experiments for stars in a Plummer potential show excellent performances in regions in the inner and outer parts of the gravity field, that is, where the motion of stars is well approximated by isochrone trajectories (with perturbations of order $10^{-3}$ or less). For highly elongated orbits that cross the characteristic length of the Plummer potential, the performance  is equivalent to that of previous methods.}
  {The splitting scheme presented here is a good alternative to previous methods: it performs at least as well, and up to orders of magnitude better, depending on the dynamical regime of the star.}


  \maketitle

\section{Introduction}\label{sec:intro}

Numerical integration of gravitational systems is a crucial component in solving many astrophysics problems. According to~\cite{Pouliasis-DiMatteo-Haywood_2017}, `The study of the orbits of stars and stellar systems, like globular and open clusters in the Milky Way, is essential to understand the properties of the different Galactic stellar populations (thin and thick discs, stellar halo) and their mode of formation'. For instance,~\cite{Khoperskov-DiMatteo-Haywood-Gomez-Snaith_2020} propose a dynamical explanation for the presence of metal-rich stars in the outer Galactic disc. But this is not the only relevant context: for example, the galactic influences to which trans-Neptunian objects in the Oort Cloud have been subjected since the formation of the Solar System --- like passing stars or galactic tides~\citep[see][for a review]{Saillenfest_2020} --- depend on the trajectory of the Sun within the Milky Way~\citep{Kaib-Roskar-Quinn_2011,MartinezBarbosa-Jilkov-PortegiesZwart-Brown_2017}. The Sun’s motion through the Milky Way may also be correlated with geological cyclic sedimentation over long timescales~\citep{Boulila_2018}. These examples, though not exhaustive, underscore the importance of integrating the trajectories of stars within the Galaxy. 

In stellar dynamics, the large number of particles makes numerical resolution extremely demanding in terms of computational performance. Fortunately, in relatively dilute regimes such as galaxies, the gravitational force acting on a star can be approximated as arising from a smooth density distribution rather than a discrete collection of mass points~\citep{Binney-Scott_2008}. At the same time, the exact form of the galactic potential remains unknown and is the subject of active research (see e.g. the discussions by~\citealp{Pouliasis-DiMatteo-Haywood_2017}). As such, `the study of individual orbits in various galactic potentials is an important branch of galactic dynamics'~\citep{Greiner_1987}. Depending on the model used and the initial conditions, orbits can change significantly (see e.g.~\citealp{Pichardo-Martos-Moreno_2004,Zotos_2014,Carita-Rodrigues-Puerari-Schiavo_2018}). Therefore, the exploration of possible trajectories often requires sampling thousands of initial conditions and integrating them all numerically over long timescales, as done for instance by~\cite{MartinezBarbosa-Jilkov-PortegiesZwart-Brown_2017}. For this purpose, the numerical integrator used must be fast and reliable, that is, it must show good stability properties over long timescales.

For planetary systems, much effort has been put into designing integrators that optimise computational cost while preserving the symplectic structure of the Hamiltonian (see e.g.~\citealp{Wisdom_1991,Chambers_1999,Laskar_2001,Farres-etal_2013,Rein-Tamayo_2015,Rein_2019_1,Hernandez-Dehnen_2024}). These integrators take advantage of the specific properties of planetary systems --- or any system with a similar hierarchy --- that are made of a dominant central mass (the `star') and other, much smaller bodies (the `planets'), such that most orbits follow perturbed Keplerian ellipses. These integrators therefore alternate between Keplerian drifts and velocity corrections. We refer to this methodology as `Kepler splitting'. Kepler splitting has shown its merits not only in the context of interacting planets (serial problem), but also for non-interacting test particles affected by known --- or pre-computed --- perturbations (the parallel problem, e.g. small moons, asteroids, and comets). By allowing people to use much larger time steps, computation costs are dramatically reduced.

To our knowledge, no dedicated integrators have been designed for the dynamics of stars immersed in effective galactic potentials, even though stellar orbits do exhibit some remarkable properties. The widely used \texttt{galpy} Python package~\citep{Bovy_2015} provides a handful of symplectic integrators, which include a second-order scheme (also known as `Leapfrog'), the fourth-order scheme from~\cite{Forest_1990}, and the sixth-order scheme from~\cite{Yoshida_1990}~\footnote{Only the first set of coefficients given by~\cite{Kinoshita-Yoshida-Nakai_1990}, referred to as SI6A, is implemented in \texttt{galpy}.}. These schemes are multi-purpose integrators in the sense that they do not take advantage of any specific property of stellar trajectories. Their principle is to alternate between linear drifts and velocity kicks --- implicitly assuming that the trajectories followed are perturbed rectilinear motions. We refer to this method as `kinetic splitting'.

Besides the limitations of kinetic splitting, the fourth-~and sixth-order schemes implemented in \texttt{galpy} include large negative sub-steps, which results in `not very good stability properties for large step sizes' (\citealp{Laskar_2001}, Sect.~1, paragraph~1; see also the discussions by~\citealp{Blanes-etal_2013} and~\citealp{Farres-etal_2013}). As a result, `[there are regimes where], at an equivalent cost, the leapfrog integrators become more effective' (\citealp{Laskar_2001}, Sect.~3, last paragraph). \cite{Pascale-Nipoti-Ciotti_2022} even conclude that in some cases, the fourth-order Runge-Kutta (RK4) method is more stable than the integrator from~\cite{Forest_1990}, despite RK4 being not symplectic. This probably explains why the integrators generally used in stellar dynamics are of low order, such as the Leapfrog (see e.g.~\citealp{Pouliasis-DiMatteo-Haywood_2017,Ferrone-etal_2023}), or `Runge-Kutta' (see e.g.~\citealp{Smirnova_2015,Boulila_2018,Sharina-Ryabova-Maricheva-Gorban_2018,Pascale-Nipoti-Ciotti_2022,Baba-Tsujimoto-Saitoh_2024,Khalil_2024}). Some integrators are symplectic, while others are not\footnote{The question whether symplectic integrators are advantageous in the context of galactic dynamics is a recurring debate. We consider they are (e.g. because systematic variations in energy would irremediably bias statistics computed on many trajectories); however, taking part in this debate is not our aim here.}.

The choice of low-order integrators in galactic dynamics also comes from the fact that using a smoothed galactic potential is itself an approximation that neglects the effect of close encounters between stars. This approximation is good for dilute systems, such as field stars in galaxies\footnote{If needed in this particular case, close encounters between stars can be modelled independently as small stochastic impulses.}, but poor for dense stellar clusters in the process of relaxation. Hence, in this context, highly accurate numerical integrations are not needed: instead, general properties of stellar dynamics are computed from an ensemble of many trajectories, whose statistics do not critically depend on the accuracy of each individual orbit (see e.g.~\citealp{PortegiesZwart-Boekholt_2014,Hernandez-etal_2020} for N-body systems). Yet, to avoid producing biased statistics, general properties of the dynamical system must be correctly reproduced by numerical integrations --- even at a reduced precision. For instance, we do not want conserved quantities (e.g. the total energy) to irreversibly drift over time away from those of the original system. To this aim, irrespective of the integrator chosen, reasonably small integration time steps must be used. When aiming for a given level of energy conservation, the question of how large the time steps can possibly be depends on the integration method --- in particular, the choice of splitting.

We investigated whether we can take advantage of the very good properties of splitting symplectic integrators in the context of stellar and galactic dynamics using an appropriate splitting of the Hamiltonian function. The goal of this new splitting is to decrease computation costs, in particular when many thousands of trajectories are needed.

Contrary to planetary --- or asteroid --- dynamics, there is no obvious choice of splitting, because trajectories in most smoothed potentials used for galaxies (e.g.~\citealp{Plummer_1911}) cannot be solved analytically. This contrasts with the dominant Keplerian potential found in planetary systems. However, one galactic potential does admit analytical solutions: Hénon’s isochrone potential~\citep{Henon_1959_1,Henon_1959_2}. Our idea is to use this potential as the basis of a new splitting scheme, which would alternate between isochrone drifts and velocity kicks. We call it `isochrone splitting'. As shown below, isochrone splitting can be seen as a generalisation of Kepler splitting. Our aim here is to use it in the context of non-collisional stellar trajectories in galactic potentials.

Even though generic galactic potentials are not isochrone, the Hénon isochrone potential does generate rosette-shaped trajectories resembling those of many stars in the plane of the Milky Way (see e.g.~\citealp{Dinescu-Girard-vanAltena_1999,Pouliasis-DiMatteo-Haywood_2017,Boulila_2018,PerezVillegas-Rossi-Ortolani-Casotto-Barbuy-Bica_2018,Ibata-Malhan-Martin_2019,Ferrone-etal_2023}). Besides, the isochrone potential is guaranteed to be a very efficient splitting scheme when we consider the orbits of distant stars in any centrally condensed potential. This property comes from the fact that all centrally condensed potentials (including the Hénon isochrone) tend to the Kepler potential at large distances. At small radii, near the centre of the mass distribution, the isochrone potential tends to a harmonic potential, which is a property shared by most potentials used in galactic dynamics (e.g.~\citealp{Plummer_1911},~\citealp{Miyamoto-Nagai_1975}, the Halo component of~\citealp{Paczynski_1990}, etc.). These characteristics of the isochrone potential make us confident that, at least in specific regimes, the isochrone potential should offer a very good kernel for a symplectic splitting.

As a first application, we present the isochrone splitting scheme for integrating the motion of a star in a Plummer potential~\citep{Plummer_1911}. Even though the Plummer potential was initially introduced to model globular clusters, it is now a fundamental building block of smoothed galactic potentials, used to compute non-collisional stellar trajectories within galaxies (see e.g.~\citealp{Allen-Santillan_1991} and other references in this section). The Plummer potential is generally used to model galactic bulges; for stars orbiting in the disc plane, the Miyamoto-Nagai potential used to model galactic discs also reduces to the Plummer potential~\citep{Miyamoto-Nagai_1975}. The omnipresence of the Plummer potential in galactic models and its simplicity motivate its use here as a testbed for the isochrone splitting scheme.

In Sect.~\ref{sec:sympinteg} we recall the basics of splitting symplectic integrators. In Sect.~\ref{sec:IsochroneOrbit} we present the analytical solutions of the isochrone dynamics needed by our integration scheme. In Sect.~\ref{sec:IsoSymp} we outline how to use these analytical solutions to build a new splitting scheme. In Sect.~\ref{sec:Plummer} we determine how to maximise the performances of the isochrone splitting scheme in the context of a particle orbiting in a Plummer potential. In Sect.~\ref{sec:perf} we illustrate our analytical results for a star orbiting in a Plummer globular cluster and compare them to those of other integration methods. Finally, we discuss our choice of integration parameters in Sect.~\ref{sec:discussion} and conclude in Sect.~\ref{sec:conclusion}.

\section{Splitting symplectic integrators}
\label{sec:sympinteg}
Throughout the article, we use the following symbol: $\mathrm{d}_{X}$ as the total derivative operator with respect to the variable $X$.

\subsection{Basics of symplectic integrators}
\label{sec:sympinteg_basics}
Let $\mathcal{H}(\mathbf{x})$ be an autonomous Hamiltonian with $m$ degrees of freedom, where $\mathbf{x}=(\mathbf{q},\mathbf{p})\in\mathbb{R}^{2m}$ is the state vector, with $\mathbf{q}\in\mathbb{R}^{m}$ denoting the generalised coordinates and $\mathbf{p}\in\mathbb{R}^{m}$ the conjugate momenta. The time evolution of the state vector is governed by Hamilton’s equation, which can be expressed as
\begin{equation}
    \label{eq:evox}
    \mathrm{d}_t\mathbf{x}=\left\{\mathbf{x},\mathcal{H}\right\}=\mathcal{L}_{\mathcal{H}}\mathbf{x}\,,
\end{equation}
where $\mathcal{L}_{\mathcal{H}}\vcentcolon=\left\{~.~,\mathcal{H}\right\}$ is the Lie derivative\footnote{We used the convention for the Poisson bracket: $\left\{f,g\right\}=\partial_{\mathbf{q}}f\cdot\partial_{\mathbf{p}}g-\partial_{\mathbf{p}}f\cdot\partial_{\mathbf{q}}g$.} along the flow of $\mathcal{H}$. The flow of $\mathcal{H}$ is defined as the map
\begin{equation}
    \label{eq:flowH}
    \phi_{t}^{\mathcal{H}}:\mathbf{x}_0\mapsto\mathbf{x}(t)\,,
\end{equation}
which associates with every point $\mathbf{x}_{0}$ in the phase space the position $\mathbf{x}$ of the system at time $t$, starting from the initial condition $\mathbf{x}_0$ at time $t=0$. The formal solution of Eq.~\eqref{eq:evox} is
\begin{equation}
    \label{eq:solx}
    \phi_{t}^{\mathcal{H}}(\mathbf{x}_0)=e^{t\mathcal{L}_{\mathcal{H}}}\phi_{0}^{\mathcal{H}}(\mathbf{x}_0)\,.
\end{equation}
In general, analytical solutions of the equations of motion do not exist, so we need to integrate Eq.~\eqref{eq:evox} numerically. In principle, any standard non-symplectic integrator can be used for this purpose (such as the methods of Euler or Runge-Kutta; see e.g.~\citealp{Hairer-Wanner-Lubich_2006}). However, these methods provide an approximate solution to an exact system of equations, with no guarantee that fundamental properties of the original system will be preserved --- such as the conservation of energy, the time reversibility, or the area-preserving nature of the flow in the phase space. Therefore, these integrators inevitably lead, over time, to a drift in total energy\footnote{What we call `energy drift' is an irreversible variation in energy produced by the non-zero error terms inherent to the integration method. This variation can be made to remain below machine precision (in which case the remaining variations in energy are only due to numerical round-off errors), but this requires using a small enough time step.}, which distorts the dynamical structure of the system. In Hamiltonian mechanics, although these types of integrators can be used, there is a way to bypass this problem by reversing the approach: instead of seeking an approximate solution to an exact system of equations, one seeks the exact solution of an approximate Hamiltonian system of equations. This type of integrator is called a symplectic integrator (see e.g.~\citealp{Yoshida_1993,Hairer-Wanner-Lubich_2006} for a comprehensive review). The solution obtained tends to oscillate around the true solution, with no irreversible drift in energy away from the real system. As such, the qualitative behaviour of the system remains correct even over long-term integrations. For accurate results, however, the oscillations artificially introduced by the symplectic integration scheme need to be minimised as much as possible.

The splitting method is the simplest kind of symplectic integrator to implement: it approximates the operator $e^{t\mathcal{L}_{\mathcal{H}}}$ in Eq.~\eqref{eq:solx}. We consider the case in which $\mathcal{H}(\mathbf{x})$ can be split into two individually integrable parts:
\begin{equation}
    \label{eq:splitH}
    \begin{aligned}
        \mathcal{H}=\mathcal{H}_{\mathrm{A}}+\mathcal{H}_{\mathrm{B}}\,.
    \end{aligned}
\end{equation}
A symplectic splitting scheme with $N$ steps is obtained by approximating $\mathcal{H}$ by a Hamiltonian $\mathcal{K}$ for which the flow $\phi^\mathcal{K}_{\delta t}$ is exactly given by
\begin{equation}
    \label{eq:expSymplec}
     \mathcal{S}_N=\prod_{i=1}^{N} e^{c_{i}\delta t\mathcal{L}_{\mathcal{H}_{\mathrm{A}}}}e^{d_{i}\delta t\mathcal{L}_{\mathcal{H}_{\mathrm{B}}}}\,,
\end{equation}
where $(c_i,d_i)\in\mathbb{R}^{2}$. Therefore, performing one time step $\delta t$ with Hamiltonian $\mathcal{K}$ simply amounts to alternate sub-steps with the (analytical) propagation of $\mathcal{H}_{\mathrm{A}}$ and $\mathcal{H}_{\mathrm{B}}$. Moreover, it is symmetric if
\begin{equation}
    \label{eq:symflowH}
    \begin{aligned}
        \mathcal{S}_{N}(\delta t)\mathcal{S}_{N}(-\delta t)=\mathcal{S}_{N}(-\delta t)\mathcal{S}_{N}(\delta t)=\mathds{1}\,,
    \end{aligned}
\end{equation}
which means that it has the exact time reversibility. In this case, the Hamiltonian $\mathcal{K}$ is necessarily even with respect to $\delta t$.

In practice, the coefficients $c_{i}$ and $d_{i}$ are chosen such that the difference $|\mathcal{H}-\mathcal{K}|$ is as small as possible. A symplectic integrator is said to be of `order $p$' if $\mathcal{H}=\mathcal{K}+\mathcal{O}(\delta t^p)$. For instance, the Leapfrog integration scheme,
\begin{equation}
    \label{eq:Oleapfrog}
    \begin{aligned}
        \mathcal{S}_{2}=e^{\frac{\delta t}{2}\mathcal{L}_{\mathcal{H}_{\mathrm{A}}}}e^{\delta t\mathcal{L}_{\mathcal{H}_{\mathrm{B}}}}e^{\frac{\delta t}{2}\mathcal{L}_{\mathcal{H}_{\mathrm{A}}}}\,,
    \end{aligned}
\end{equation}
is a symmetric symplectic scheme of order~$2$.

Very efficient symplectic schemes can be obtained if the splitting of $\mathcal{H}$ in Eq.~\eqref{eq:splitH} is hierarchical, that is, if we can write $\mathcal{H}_{\mathrm{A}} = \mathcal{A}$ and $\mathcal{H}_{\mathrm{B}} = \varepsilon\mathcal{B}$, where $\varepsilon\ll 1$. For instance, the symmetric integrators $\mathcal{SABA}_n$ and $\mathcal{SABAC}_n$ of~\cite{Laskar_2001} have errors $\mathcal{H}-\mathcal{K}=\mathcal{O}(\varepsilon\delta t^{2n}) + \mathcal{O}(\varepsilon^2\delta t^2)$, and $\mathcal{O}(\varepsilon\delta t^{2n}) + \mathcal{O}(\varepsilon^2\delta t^4)$, respectively\footnote{$\mathcal{SABA}_1$ corresponds to the Leapfrog integration scheme.}. The fact that powers of $\varepsilon$ appear in the remainders means that if $\varepsilon$ is small, even a scheme with a small number of steps $N$ gives very good performances. For instance, the integrator $\mathcal{SABA}_3$ has~$4$ steps and it behaves as an integrator of order~$6$ if $\varepsilon$ is small enough, despite remainders being formally of order~$2$ in $\delta t$. For comparison, the integrator given by~\cite{Forest_1990} has the same number of steps as $\mathcal{SABA}_3$ but it is of order~$4$~\citep{Laskar_2001}. More general symplectic schemes have been developed by~\cite{Blanes-etal_2013}, with remainders of the form $\mathcal{H}-\mathcal{K} = \mathcal{O}(\varepsilon\delta t^{s_1}) + \mathcal{O}(\varepsilon^2\delta t^{s_2}) + \mathcal{O}(\varepsilon^3\delta t^{s_3}) + \dots$ where $\{s_1,s_2,s_3\dots\}$ are even integer coefficients. These integrators are symmetric and do not have negative coefficients $c_{i}$ and $d_{i}$; or, if they do, such coefficients are few and very small (lower than one in absolute value). According to~\cite{Laskar_2001}, this increases the global numerical stability of the integration scheme (see also~\citealp{Blanes-etal_2013,Farres-etal_2013}).

\subsection{The choice of splitting}
In our case, we consider a generic Hamiltonian function $\mathcal{H}$ in the form
\begin{equation}
    \label{eq:genericH}
    \begin{aligned}
        \mathcal{H}(\mathbf{r}, \dot{\mathbf{r}})= \frac{\|\dot{\mathbf{r}}\|^2}{2} + \Psi(\mathbf{r}) \,,
    \end{aligned}
\end{equation}
where $\Psi(\mathbf{r})$ is a potential. Any splitting scheme can be introduced by rewriting Eq.~\eqref{eq:genericH} as
\begin{equation}
    \label{eq:Ham}
    \begin{aligned}
        \mathcal{H}(\mathbf{r}, \dot{\mathbf{r}})=\mathcal{A}(\mathbf{r}, \dot{\mathbf{r}})+\varepsilon\mathcal{B}(\mathbf{r})\,,
    \end{aligned}
\end{equation}
where
\begin{align}
    \label{eq:A-HAM}
    \mathcal{A}(\mathbf{r}, \dot{\mathbf{r}}) &= \frac{\|\dot{\mathbf{r}}\|^2}{2} + \Phi(\mathbf{r})\,,\\
    \label{eq:eB-HAM}
    \varepsilon\mathcal{B}(\mathbf{r}) &= \Psi(\mathbf{r}) - \Phi(\mathbf{r})\,,
\end{align}
and $\Phi(\mathbf{r})$ is an arbitrary potential. In practice, we choose $\Phi(\mathbf{r})$ such that the dynamics driven by Hamiltonian function $\mathcal{A}$ can be solved analytically.

The most simple choice of splitting is to set $\Phi(\mathbf{r})=0$; this corresponds to the classic kinetic splitting, used for instance in the popular Leapfrog symplectic integrator~\citep{Ruth_1983}. Yet, depending on the problem under study, other choices can be much more judicious. In particular, if the chosen potential $\Phi(\mathbf{r})$ is close to $\Psi(\mathbf{r})$, then $\mathcal{A}$ becomes a dominant integrable part and $\varepsilon\mathcal{B}$ is a small perturbation. We can then solve for the dynamics using symplectic integrators for perturbed Hamiltonian systems (see e.g.~\citealp{McLachlan_1995,Laskar_2001,Blanes-etal_2013}). Depending on the magnitude $\varepsilon$ of the perturbation, these more judicious splitting schemes can be many orders of magnitude more efficient than kinetic splitting (see Sect.~\ref{sec:sympinteg_basics}). This is the case in planetary systems, for which the motion of planets is a perturbed Keplerian orbit; in that case, we choose $\Phi(\mathbf{r}) = -\mu/\|\mathbf{r}\|$, and the parameter $\mu$ and coordinates $\mathbf{r}$ are selected to give the best performances (see e.g.~\citealp{Farres-etal_2013,Hernandez_2017,Rein_2019_1}); this corresponds to Kepler splitting.

When dealing with the motion of a star in a stellar cluster or a galaxy, there is no obvious choice of splitting potential, because $\Psi(\mathbf{r})$ can be composed of many different parts, which are generally not integrable and have no particular hierarchy (see e.g.~\citealp{Paczynski_1990,Gardner-Nurm-Flynn-Mikkola_2011,Pouliasis-DiMatteo-Haywood_2017}). Yet, one galactic potential has very interesting properties: the H{\'e}non isochrone potential (see e.g.~\citealp{Simon-Petit_2018,Ramond-Perez_2020}). As the isochrone dynamics is analytically integrable and the Hénon isochrone potential does reproduce realistic distributions of mass in stellar dynamics (e.g. in globular clusters; see~\citealp{Henon_1959_1}), we investigated whether it can be used as an efficient splitting scheme.

\section{Analytical solution of the isochrone dynamics}
\label{sec:IsochroneOrbit}
A potential $\Phi(r)$, where $r$ is the radial coordinate, is called isochrone if all non-radial bounded orbits have a radial period that does not depend on the angular momentum. As a theorem,~\cite{Ramond-Perez_2021} have shown that this definition is fulfilled if and only if the curve $y=\mathcal{Y}(x)$, where $x=2r^2$ and $\mathcal{Y}(x)=x\Phi(r(x))$, depicts a (convex arc of) parabola in the $xOy$ plane. The Kepler potential is an example of the application of this theorem. The family of isochrone potentials can be divided into several groups based on the orientation of the isochrone parabola $\mathcal{Y}(x)$ in the $xOy$ plane and the number of intersections of the parabola with the $y$-axis~\citep{Simon-Petit_2018}. The Hénon isochrone potential, often referred to simply as the `isochrone potential', is one of this family and is widely used in various contexts in galactic dynamics. In fact, `...the isochrone potential is guaranteed a role in dynamical astronomy because it is the most general potential in which closed-form expressions for angle-action coordinates are available'~\citep{Binney_2014}. Initially introduced by~\cite{Henon_1959_1} to reproduce the mass distribution of globular clusters, this potential accurately describes any self-gravitating system whose isochrone characteristics have not been erased by dynamical perturbations during its evolutionary process~\citep{Simon-Petit_2019}. Our goal here is to use this potential as part of a splitting scheme to integrate the motion of stars immersed in effective galactic potentials.

To implement our proposed splitting scheme, the first step is to express the explicit analytical solution of the two parts, $\mathcal{A}$ and $\varepsilon\mathcal{B}$, taken individually. The integration of $\varepsilon\mathcal{B}$ is trivial because it only depends on the position. The integration of $\mathcal{A}$ requires more work.

Let $\mathcal{A}$ be the Hamiltonian of the system made of a unit mass test particle orbiting in Hénon's isochrone potential $\Phi(\mathbf{r})$. In terms of position $\mathbf{r}$ and velocity $\dot{\mathbf{r}}$, the Hamiltonian reads
\begin{equation}
    \label{eq:Ham_A}
    \begin{aligned}
        \mathcal{A}(\mathbf{r}, \dot{\mathbf{r}})=\frac{\|\dot{\mathbf{r}}\|^2}{2}-\frac{\mu}{b+\sqrt{\|\mathbf{r}\|^2+b^2}}
        \,,
    \end{aligned}
\end{equation}
where $\left(\mu,b\right)\in\left(\mathbb{R}_+\right)^2$. 
The isochrone parameter $b$ is a characteristic length that quantifies the way the mass is distributed in the physical system, and it is related to the half-mass radius --- the radius of the sphere that contains half the total mass --- as follows:
\begin{equation*}
    \begin{aligned}
        r_{1/2}=\left(8+\sum_{\pm}\sqrt[3]{1592\pm81\sqrt{327}}\right)\frac{b}{9}
        \approx 3 b
        \,.
    \end{aligned}
\end{equation*}
In this equation, the summation is made over two terms, for which the `$\pm$' symbol is replaced by `$+$' and by `$-$'.
The potential is Keplerian if all the mass is concentrated at a single point ($b=0$). More generally, when $\|\mathbf{r}\| \gg b$, the potential in Eq.~\eqref{eq:Ham_A} felt by the particle is Keplerian, and it is harmonic when $\|\mathbf{r}\| \ll b$.

The analytical solution to the equations of motion for a bound orbit generated by any isochrone potential is provided by~\cite{Ramond-Perez_2021}. However, to use Hénon's potential as a splitting scheme, all possible types of isochrone trajectories must be implemented, including unbound trajectories. Inspired by~\cite{Ramond-Perez_2021}, we derived here the solutions for all possible kinds of trajectories in a formalism adapted to our goals.

We note that there is a vast literature dedicated to the Kepler problem (i.e. for $b=0$), including sets of variables that allow one to describe all kinds of trajectories at once --- bounded and unbounded (see e.g.~\citealp{Wisdom-Hernandez_2015}). The derivation of analogous variables for isochrone orbits and similar algorithmic refinements are left for future works.

As the angular momentum $\mathbf{\Lambda}=\mathbf{r}~\times~\dot{\mathbf{r}}$ is constant, the particle's motion lies on a plane. On this plane, the particle's position can be described by its radial distance $r$ and polar angle $\varphi$. We now need to express $r$ and $\varphi$ as functions of time $t$ for any kind of trajectory. From expanding the kinetic energy in the Hamiltonian in Eq.~\eqref{eq:Ham_A}, we can express the energy constant as
\begin{equation}
    \label{eq:h_of_A}
    \begin{aligned}
        h=\frac{1}{2}\left(\left(\mathrm{d}_t r\right)^2+\left(\frac{\Lambda}{r}\right)^2\right)-\frac{\mu}{b+\sqrt{r^2+b^2}}\,,
    \end{aligned}
\end{equation}
where $\Lambda=\|\mathbf{\Lambda}\|$.

\subsection{Bound orbit: Negative particle's energy}
\label{sec:Bound_orbit}
A bound orbit corresponds to a negative total energy: $h < 0$. In that case, the particle oscillates between two values of $r$ called the periapsis, $r_{\mathrm{p}}$, and the apoapsis, $r_{\mathrm{a}}$, which can be expressed as a function of $\Lambda$ and $h$.

Following Appendix~\ref{an:calDetailEQDIFF}, the time and position of the particle along its orbit are related through
\begin{align}
    t &= \int\frac{\mathrm{d}r}{\sqrt{2\left(h-\Phi(r)\right)-\left(\Lambda/r\right)^2}} \\
    \label{eq:deltat}
    &= \frac{\mathbf{r}\cdot\dot{\mathbf{r}}}{2h}-\frac{\mu}{\sqrt{-2h^{3}}}\arctan\left(\frac{-1}{\sqrt{-2h}}\mathrm{d}_{\sqrt{r^2+b^2}}\left(\mathbf{r}\cdot\dot{\mathbf{r}}\right)\right)\,.
\end{align}
The dot product between the position and the velocity is zero at periapsis and apoapsis. Therefore, the travel time from periapsis to apoapsis is
\begin{equation}
    \label{eq:tlimit}
    \begin{aligned}
       \Delta t_{r_{\mathrm{p}}\rightarrow r_{\mathrm{a}}}=\frac{\mu}{\sqrt{-2h^{3}}}\left[\lim_{s\to0^{+}}\arctan\left(\frac{\mathrm{d}_u s}{\sqrt{-2h}}\right)\right]
       \,,
    \end{aligned}
\end{equation}
where $s=\mathbf{r}\cdot\dot{\mathbf{r}}$ and
\begin{equation}
    \label{eq:udef}
    u=\sqrt{1+\left(r/b\right)^2}\,.
\end{equation}
The variables $u$ and $s$ depend on each other as
\begin{equation}
    \label{eq:sfromu}
    s = \sqrt{k_3(u^2-1)+k_2(u-1)-k_1}\,,
\end{equation}
where $k_1 = \left(\Lambda/b^2\right)^2$, $k_2=2\mu/b^3$, and $k_3=2h/b^2$. From Eq.~\eqref{eq:sfromu}, $\mathrm{d}_u s$ in Eq.~\eqref{eq:tlimit} behaves as $1/s$ when $s$ approaches $0^+$. A radial period $\tau$ corresponds to a complete cycle ($r_{\mathrm{p}} \rightarrow r_{\mathrm{a}} \rightarrow r_{\mathrm{p}}$), which gives
\begin{equation}
    \label{eq:KeplerLaw}
    \tau = 2\pi\frac{\mu}{\left|2h\right|^{1.5}}\,.
\end{equation}
This is Kepler’s third law. By analogy with the resolution of the two-body problem, we can define an eccentricity, $\epsilon,$ and an eccentric anomaly, $E,$ that obey Kepler's equation:
\begin{equation}\label{eq:KeplerEq}
    n\Delta t = E - \epsilon\sin E\,,
\end{equation}
where $n=2\pi/\tau$ is the mean motion. By identification of the two terms in the right-hand side of Eqs.~\eqref{eq:deltat} and~\eqref{eq:KeplerEq}, we set
\begin{equation*}
    \begin{aligned}
        \frac{n}{2h}\mathbf{r}\cdot\dot{\mathbf{r}}=-\epsilon\sin{E}\,,
    \end{aligned}
\end{equation*}
where $\mathbf{r}\cdot\dot{\mathbf{r}}=b^2\mathrm{d}_t u^2/2$ (see Eq.~\eqref{eq:udef}). The previous expression can be rewritten as
\begin{equation*}
    \begin{aligned}
        b^2u^2=\alpha^2\int 2\epsilon\sin{E}\left(1-\epsilon\cos{E}\right)\mathrm{d}E\,,
    \end{aligned}
\end{equation*}
where we define the semi-major axis as $\alpha=-\mu/2h$, such that $\mu=n^2\alpha^3$ (as in the two-body problem). Finally, the radial solution is
\begin{equation}
    \label{eq:r(E)}
    \begin{aligned}
        r\left(E\right)^2=\alpha^2\left(1-\epsilon\cos{E}\right)^2-b^2\,,
    \end{aligned}
\end{equation}
with the eccentricity 
\begin{equation}
    \label{eq:eccentricity_h<0}
    \begin{aligned}
        \epsilon = \sqrt{\left(1-\frac{b}{\alpha}\right)^2-\frac{\Lambda^2}{\mu\alpha}}\,.
    \end{aligned}
\end{equation}

It remains to express the polar angle $\varphi$ as a function of $E$. We obtain
\begin{equation*}
    \begin{aligned}
        \mathrm{d}_t \varphi=\frac{\Lambda}{r^2} \iff \mathrm{d}\varphi &= \frac{\Lambda}{\sqrt{\mu\alpha}}\frac{1-\epsilon\cos E}{\left(1-\epsilon\cos E\right)^2-\left(b/\alpha\right)^2}\mathrm{d}E \\
        &=\frac{\Lambda}{\sqrt{\mu\alpha}}\sum_{\pm}\frac{1-\epsilon\cos E}{\left(1\pm b/\alpha-\epsilon\cos E\right)}\mathrm{d}E\,,
    \end{aligned}
\end{equation*}
which, when integrated, gives the angular solution:
\begin{equation}
    \label{eq:phi(E)}
    \begin{aligned}
        \varphi\left(E\right) = \frac{\Lambda}{\sqrt{\mu\alpha}}\sum_\pm \frac{\arctan \left(\sqrt{\frac{1\pm (b/\alpha)+\epsilon}{1\pm (b/\alpha)-\epsilon}}\tan \left(E/2\right)\right)}{\sqrt{\left(1\pm\left(b/\alpha\right)\right)^2-\epsilon^2}}\,.
    \end{aligned}
\end{equation}Table~\ref{tab:BoundOrbitForm} shows all the possible forms of bound orbits (see Appendix~\ref{an:detailOrbit_h<0} for details).

{\renewcommand{\arraystretch}{1.3}
\begin{table*}
    \caption{Classification of isochrone bound orbits.}
    \label{tab:BoundOrbitForm}
    \centering
    \begin{tabular}[b]{lccc}
        \hline
        \multirow{2}{*}{Orbit shape} & \multirow{2}{*}{Isochrone parameter} & \multicolumn{2}{c}{Orbital parameters} \\
        \cline{3-4}
        & & Semi-major axis & Eccentricity \\
        \hline \hline
        Point & \multirow{3}{*}{$b\ge0$} & $\alpha=b$ & \multirow{2}{*}{$\epsilon=0$} \\
        \cline{1-1}\cline{3-3}
        Circle & & \multirow{4}{*}{$\alpha>b$} & \\
        \cline{1-1}\cline{4-4}
        Line segment & & & $\epsilon=1-b/\alpha$ \\
        \cline{1-2}\cline{4-4}
        Ellipse & $b=0$ & & $0<\epsilon<1$ \\
        \cline{1-2}
        Rosette & $b>0$ & & $\epsilon\neq1-b/\alpha$ \\
        \hline
    \end{tabular}
    \tablefoot{Rosette orbits consist of a succession of loops of the same size. Special cases of rosette orbits include the ellipse, circle, and line segment (the loop has zero area). The orbit shape called `Point' corresponds to a situation where the particle does not move.}
\end{table*}
}

\subsection{Unbound orbit: Positive particle's energy}
\label{sec:Unbound_orbit-h>0}
When the total energy in Eq.~\eqref{eq:h_of_A} is positive ($h>0$), the orbit is unbound and Eq.~\eqref{eq:deltat} remains valid. Similarly to the previous (see Sect.~\ref{sec:Bound_orbit}), the hyperbolic Kepler equation allows us to define a hyperbolic eccentric anomaly $H$ and an eccentricity $\epsilon$:
\begin{equation}
    \label{eq:hyperbKepler}
    n\Delta t = \epsilon\sinh H - H\,.
\end{equation}
Again, by identification of the two terms in the right-hand side of Eqs.~\eqref{eq:deltat} and~\eqref{eq:hyperbKepler}, in a manner similar to the previous section, we set
\begin{equation*}
    \begin{aligned}
        \frac{n}{2h}\mathbf{r}\cdot\dot{\mathbf{r}}=\epsilon\sinh{H}\,,
    \end{aligned}
\end{equation*}
which can be rewritten as
\begin{equation*}
    \begin{aligned}
        b^2u^2=\alpha^2\int 2\epsilon\sinh{H}(\epsilon\cosh{H}-1)~\mathrm{d}H\,,
    \end{aligned}
\end{equation*}
so the radial solution is\begin{equation}
    \label{eq:r(H)}
    \begin{aligned}
        r\left(H\right)^2=\alpha^2\left(\epsilon\cosh H -1\right)^2-b^2\,,
    \end{aligned}
\end{equation}
with the eccentricity 
\begin{equation*}
    \begin{aligned}
        \epsilon = \sqrt{\left(1+\frac{b}{\alpha}\right)^2+\frac{\Lambda^2}{\mu\alpha}}\,,
    \end{aligned}
\end{equation*}
and the semi-major axis $\alpha=\mu/2h$. As for the polar angle as a function of $H$, we obtain
\begin{equation*}
    \begin{aligned}
        \mathrm{d}\varphi &= \frac{\Lambda}{\sqrt{\mu\alpha}}\frac{\epsilon\cosh H -1}{(\epsilon\cosh H -1)^2-(b/\alpha)^2}\mathrm{d}H \\
        &= \frac{\Lambda}{\sqrt{\mu\alpha}}\sum_{\pm}\frac{\epsilon\cosh H -1}{(\epsilon\cosh H\pm b/\alpha-1)}\mathrm{d}H\,,
    \end{aligned}
\end{equation*}
which, when integrated, gives the angular solution:
\begin{equation}
    \label{eq:phi(H)}
    \begin{aligned}
        \varphi\left(H\right) = \frac{\Lambda}{\sqrt{\mu\alpha}}\sum_\pm \frac{\arctan \left(\sqrt{\frac{\epsilon + 1\pm (b/\alpha)}{\epsilon-(1\pm (b/\alpha))}}\tanh \left(H/2\right)\right)}{\sqrt{\epsilon^2-(1\pm(b/\alpha))^2}}\,.
    \end{aligned}
\end{equation}
When $H$ tends to $-\infty$ or $+\infty$, the trajectory tends to the branches of a Keplerian hyperbola.

\subsection{Unbound orbit: Zero particle's energy}
\label{sec:Unbound_orbit-h=0}
When the total energy in Eq.~\eqref{eq:h_of_A} is zero ($h=0$), the orbit is unbound, and Eq.~\eqref{eq:deltat} is undefined. However, we can use the same change of variable as in Sect.~\ref{sec:Bound_orbit}, with the relation
\begin{equation*}
    \left(\mathrm{d}_t u^2\right)^2=4s^2\,.
\end{equation*}
Following Appendix~\ref{an:calDetailEQDIFF} with $h=0$, this equation yields the polynomial
\begin{equation*}
    \label{eq:poly_s}
    \begin{aligned}
         \Tilde{s}^3+3\left(\Lambda^2+2b\mu\right)\Tilde{s}=6\mu^2t\,,
    \end{aligned}
\end{equation*}
with $\Tilde{s}=b^2s$, and which can be solved using Cardano's method. The unique real root provides the dot product between the position and velocity at the given instant: $s=f(t)$, where
\begin{equation*}
    \label{eq:rdotdr_dt-h=0}
    \begin{aligned}
        f(t)=\varsigma\left(t\right)-\frac{\Lambda^2+2b\mu}{\varsigma\left(t\right)}\,,
    \end{aligned}
\end{equation*}
and $\varsigma\left(t\right)=3^{1/3}\left(\mu^2t+\sqrt{\mu^4t^2+\frac{\left(\Lambda^2+2b\mu\right)^3}{9}}\right)^{1/3}$. The radial solution in this case is given by
\begin{equation}
    \label{eq:r(rtdr_dt)}
    \begin{aligned}
    r^2=\left(\frac{\Lambda^2+f(t)^2}{2\mu}+b\right)^2-b^2\,.
    \end{aligned}
\end{equation}

To obtain the temporal evolution of the polar angle $\varphi$, one needs to take the limit of Eq.~\eqref{eq:phi(E)} --- or Eq.~\eqref{eq:phi(H)} --- as the total energy, $h$, approaches zero. As detailed in Appendix~\ref{an:polaran}, we obtain
\begin{equation}
    \label{eq:phi(rdr_dt)}
    \begin{aligned}
        \varphi_{_{h \to 0}}=\frac{\Lambda}{\sqrt{\Lambda^2+4b\mu}}\arctan{\left(\frac{f(t)}{\sqrt{\Lambda^2+4b\mu}}\right)}+\arctan{\left(\frac{f(t)}{\Lambda}\right)}\,.
    \end{aligned}
\end{equation}
When $t$ tends to $-\infty$ or $+\infty$, the trajectory tends to the branches of a Keplerian parabola.

\section{The isochrone splitting scheme}
\label{sec:IsoSymp}
\begin{figure*}
    \includegraphics[width=\textwidth]{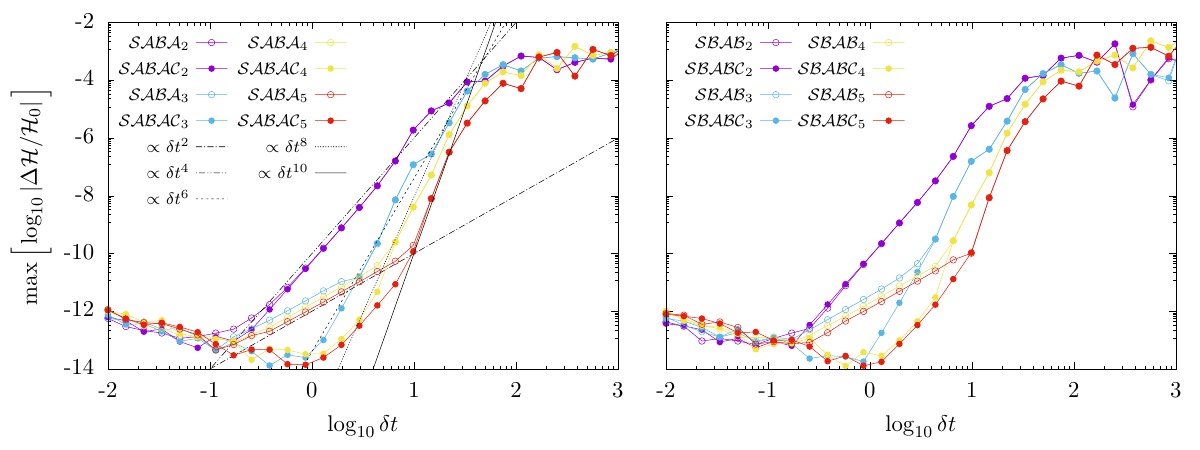}
    \caption{Maximum relative variation in the Hamiltonian in Eq.~\eqref{eq:Htest} as a function of the time step used. We integrate a slightly perturbed isochrone potential where $\varepsilon = 10^{-4}$. The integration is conducted over two radial periods ($\Delta t=2\tau\approx 239$) of an orbit with unperturbed isochrone semi-major axis $\alpha\approx 7.12$ and eccentricity $\epsilon\approx 4.21\times10^{-1}$. Black lines show several power laws for comparison.}
    \label{fig:KepC}
\end{figure*}
Now that all solutions of the isochrone dynamics are available, we can use them to build a symplectic splitting scheme, as described in Sect.~\ref{sec:sympinteg}. For sub-steps with Hamiltonian $\mathcal{A}$ (i.e. an unperturbed isochrone motion), we used the formulas described in Sect.~\ref{sec:IsochroneOrbit} to propagate any position and velocity $(\mathbf{r},\dot{\mathbf{r}})$ to their new value. We followed the standard methodology used for Kepler solvers (see e.g.~\citealp{Rein-Tamayo_2015}). Yet, we did not find isochrone analogous versions of the Gauss functions~$f$~and~$g$~\citep{Danby_1988}, which means that our algorithm needs to compute the true anomaly $\varphi$ explicitly. While doing so, we had to take special care that the value returned by the two arc-tangents (see the formulas in Sect.~\ref{sec:IsochroneOrbit}) lie in the correct quadrant of the trigonometric circle. The outline of an algorithm to compute the isochrone sub-steps is given in Appendix~\ref{sec:algoiso}.

To validate our analytical computations and numerical implementations, we first applied the isochrone splitting scheme to a slightly perturbed isochrone dynamics. We chose a dimensionless Hamiltonian function of the form
\begin{equation}
     \label{eq:Htest}
     \mathcal{H}(\mathbf{r}, \dot{\mathbf{r}}) = \frac{1}{2}\|\dot{\mathbf{r}}\|^2 -\frac{\eta}{\kappa + \sqrt{\|\mathbf{r}\|^2 + \kappa^2}} - \varepsilon\,\frac{\eta}{\|\mathbf{r}\|}\,,
\end{equation}
where $\eta=1$, $\kappa=1$ and $\varepsilon$ is a small quantity. To apply the isochrone splitting scheme, we split the Hamiltonian function as in Eqs.~\eqref{eq:A-HAM} and~\eqref{eq:eB-HAM}, in which $\Psi(r) = -\eta/(\kappa + \sqrt{r^2 + \kappa^2}) - \varepsilon\eta/r$ and $\Phi(r) = -\mu/(b + \sqrt{r^2 + b^2})$, where we chose $\mu=\eta$ and $b=\kappa$.

\begin{figure}
    \includegraphics[width=0.8\columnwidth]{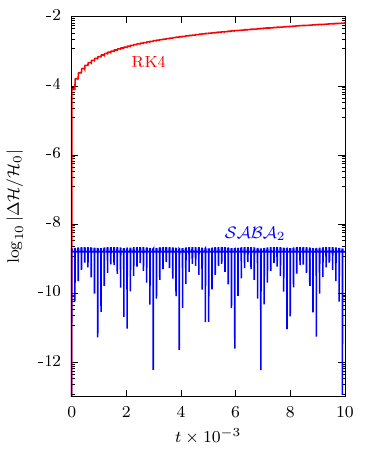}
    \caption{Relative error in total energy as a function of time for the same orbit as in Fig.~\ref{fig:KepC}. The behaviour of two fourth-order integrators is shown: the Runge-Kutta method and the $\mathcal{SABA}_2$ method using the isochrone splitting (see the labels). A constant time step $\delta t=2.5$ is used for both integrators.}
    \label{fig:SABA2vsRK4}
\end{figure}

Figure~\ref{fig:KepC} illustrates the results obtained for the $\mathcal{SABA}_n$, $\mathcal{SBAB}_n$, $\mathcal{SABAC}_n$, and $\mathcal{SBABC}_n$ families of integrators, where $n = 2$ to~$5$. We note $\mathcal{H}_0$ the Hamiltonian at the initial time and $\Delta\mathcal{H}$ its instantaneous variation relative to $\mathcal{H}_0$ at time~$t$. As introduced in Sect.~\ref{sec:sympinteg}, the integrators of~\cite{Laskar_2001} have errors of order $\mathcal{O}(\varepsilon\delta t^{2n}) + \mathcal{O}(\varepsilon^2\delta t^2)$ for $\mathcal{SABA}_n$ and $\mathcal{SBAB}_n$, and $\mathcal{O}(\varepsilon\delta t^{2n}) + \mathcal{O}(\varepsilon^2\delta t^4)$ for $\mathcal{SABAC}_n$ and $\mathcal{SBABC}_n$, where $\delta t$ is the integration step. As expected, Fig.~\ref{fig:KepC} shows that depending on the value of $\delta t$, either the first or the second error term dominates (see the curve slopes). This validates our implementation of the isochrone splitting scheme. As shown in Fig.~\ref{fig:SABA2vsRK4}, the use of a symplectic integrator (e.g. $\mathcal{SABA}_2$) with the isochrone splitting produces bounded energy variations, and no irreversible drift of energy over long timescales. In contrast, the RK4 non-symplectic integrator produces an energy drift that is visible within the first radial period.

To apply our splitting scheme to a generic potential $\Psi(\mathbf{r})$, which may have an expression that greatly differs from the isochrone potential, we must find a generic way of choosing the parameters $\mu$ and $b$ of the isochrone potential $\Phi(\mathbf{r})$ used for the splitting. To maximise the efficiency of our symplectic splitting scheme (see Sect.~\ref{sec:sympinteg}), we need to minimise the size of the perturbation $\varepsilon\mathcal{B}$ as much as possible all along the trajectory. A possible choice is to set
\begin{equation}
    \label{eq:mub-generalcase}
    \left\{
    \begin{aligned}
        \Big(\varepsilon\mathcal{B}\Big)_{r=q}&=0 \\
        \Big(\mathrm{d}_r\varepsilon\mathcal{B}\Big)_{r=q}&=0
    \end{aligned}
    \right.
    ~~~~\iff~~~~
    \left\{
    \begin{aligned}
        b&=\left(\frac{r\left(1+r\frac{\mathrm{d}_r\Psi}{\Psi}\right)}{\sqrt{1-\left(1+r\frac{\mathrm{d}_r\Psi}{\Psi}\right)^2}}\right)_{r=q} \\
        \mu&=-\left(\left(\sqrt{r^2+b^2}+b\right)\Psi\right)_{r=q}
    \end{aligned}
    \right. 
    \,,
\end{equation}
where $q$ is a given distance (e.g. the initial radial distance of the particle). This choice for the parameters $\mu$ and $b$ implies that if $\varepsilon\mathcal{B}$ solely depends on the radial distance $r$, the integration is exact whenever the star is located at a distance $r=q$. This parameter choice is studied in Sect.~\ref{sec:Plummer} for a Plummer potential. Equation~\eqref{eq:mub-generalcase} shows that this way of choosing the parameters $\mu$ and $b$ for the isochrone splitting function $\Phi(\mathbf{r})$ cannot be applied to any potential $\Psi(\mathbf{r})$; the following property must be satisfied:
\begin{equation}
    \begin{aligned}
        -1\leq \left(r\frac{\mathrm{d}_r\Psi}{\Psi}\right)_{r=q}<0\,.
    \end{aligned}
\end{equation}
If this condition is not verified, then the parameters $\mu$ and $b$ must be chosen in a different way (see Sect.~\ref{sec:discussion} for more details). 

We note that, if we set $b=0$, we retrieve the Kepler splitting scheme widely used in celestial mechanics; in that case, $\mu$ is generally chosen as the gravitational parameter of the central body (see e.g.~\citealp{Farres-etal_2013,Hernandez_2017,Rein_2019_1}). Here, the addition of a second parameter --- the length scale, $b$ --- gives us more freedom to adjust the integration scheme and adapt it to the particular problem under study.

\section{Application: Motion of a star in a Plummer potential}
\label{sec:Plummer}
As a first experimentation of our new splitting scheme, we chose to solve for the motion of a star in a Plummer potential~\citep{Plummer_1911}. The Plummer potential is the most simple galactic potential for which no closed-form analytic solution exists. The Plummer potential is an essential component of most galactic potentials (see e.g.~\citealp{Paczynski_1990,Allen-Santillan_1991,Pouliasis-DiMatteo-Haywood_2017}). It is therefore an unavoidable testbed for integrating the motion of stars in galaxies. This potential is expressed as
\begin{equation}
    \label{eq:plummer}
    \begin{aligned}
        \Psi(\mathbf{r})=-\frac{\eta}{\sqrt{\|\mathbf{r}\|^2+\kappa^2}}\,,
    \end{aligned}
\end{equation}
where $\left(\eta,\kappa\right)\in\left(\mathbb{R}_+\right)^2$. Even though there is no closed-form solution for the motion of the star in this potential, we know that the motion is planar. Besides, the radial distance of the star either oscillates between the periapsis distance, $r_{\mathrm{p}}$, and the apoapsis distance, $r_{\mathrm{a}}$ (bounded orbit), or it goes to infinity before and after reaching the star's periapsis, $r_{\mathrm{p}}$ (unbound orbit).

To apply our new splitting scheme and integrate the motion of the star, we divided the Hamiltonian function of the star as in Eq.~\eqref{eq:Ham} such that the perturbation is
\begin{equation}
    \label{eq:Ham_epsB}
    \begin{aligned}
        \varepsilon\mathcal{B}(\mathbf{r})=\frac{\mu}{b+\sqrt{\|\mathbf{r}\|^2+b^2}}-\frac{\eta}{\sqrt{\|\mathbf{r}\|^2+\kappa^2}}\,.
    \end{aligned}
\end{equation}
If $\varepsilon\mathcal{B}$ is zero, then the star follows an isochrone orbit, and the solutions to the equations of motion are exact (see Sect.~\ref{sec:IsoSymp}). Following Eq.~\eqref{eq:mub-generalcase}, the isochrone mass parameter $\mu$ and scale radius $b$ used for the splitting can be chosen as follows:
\begin{equation}
    \label{eq:mub}
    \left\{
    \begin{aligned}
        b&=\frac{\kappa}{\sqrt{2+\left(q/\kappa\right)^2}}\\
        \mu&=\eta\sqrt{\frac{2+\left(q/\kappa\right)^2}{1+\left(q/\kappa\right)^2}}
    \end{aligned}
    \right.
    \,,
\end{equation}
where $q$ is an arbitrary value of $r$ taken along the motion of the star\footnote{In practice, as we want $|\varepsilon\mathcal{B}|$ to be as small as possible, we must take care of the dramatic cancellation errors generated by the difference in Eq.~\eqref{eq:Ham_epsB} by implementing the equation in a cancellation-safe way (see Appendix~\ref{an:cancellation}).}.

The optimal choice of the distance $q$ chosen to compute $\mu$ and $b$ in Eq.~\eqref{eq:mub} is not trivial. With no lack of generality, we used normalised distances $\tilde{r}=r/\kappa$ and $\tilde{q}=q/\kappa$ and the normalised time $\mathrm{d}\tilde{t}=(\eta/\kappa)\mathrm{d}t$, which is equivalent to multiplying the total Hamiltonian function by the constant factor $\kappa/\eta$. In what follows, we drop the tilde symbol `$\sim$' for readability. In this system of units, Eq.~\eqref{eq:Ham_epsB} can be expressed as\begin{equation}
    \label{eq:eB0}
    \varepsilon\mathcal{B}_q(r)
    = \frac{q^2 + 2}{\sqrt{q^2+1}\left(1+\sqrt{1+r^2\left(q^2+2\right)}\right)}-\frac{1}{\sqrt{1+r^2}}\,.
\end{equation}
We can prove that the function $\varepsilon\mathcal{B}_q(r)$ is positive for any value of the variable $r$ and parameter $q$ (see Appendix~\ref{an:Des_epsB>0}). To simplify the notations, we use below the index `$\mathrm{p}$' when $q=r_\mathrm{p}$ and the index `$\mathrm{a}$' when $q=r_\mathrm{a}$.

\begin{figure}
    \includegraphics[width=\columnwidth]{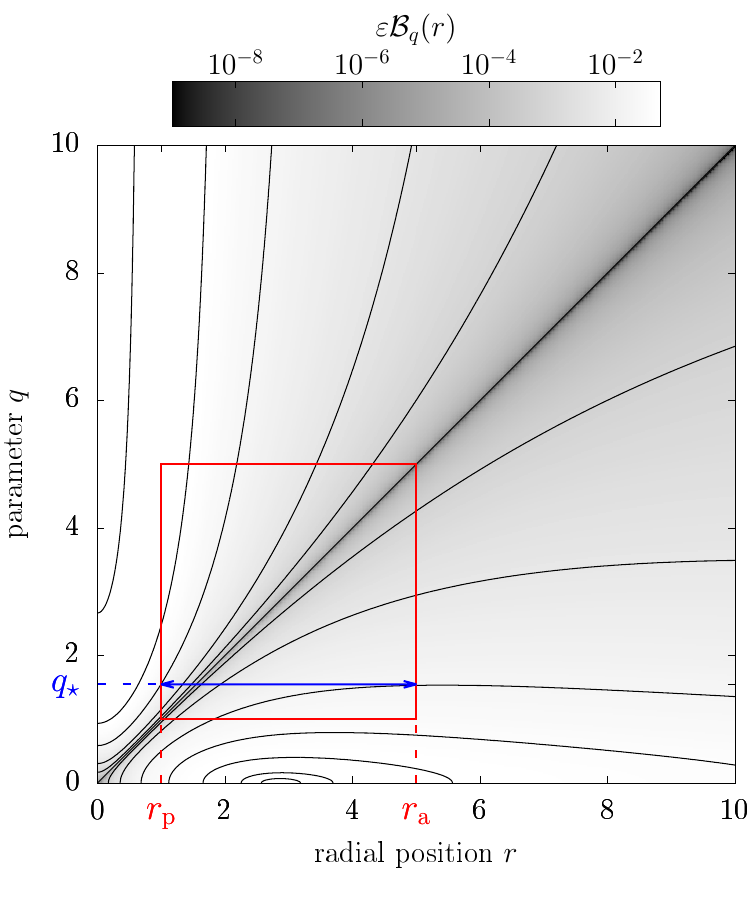}
    \caption{Value of the Hamiltonian $\varepsilon\mathcal{B}_q(r)$ in Eq.~\eqref{eq:eB0} as a function of the parametrisation point, $q,$ and the star's instantaneous position, $r$. The value of $\varepsilon\mathcal{B}_q(r)$ is shown by the colour shades and black level curves. By definition of $q$, the value of $\varepsilon\mathcal{B}_q(r)$ is zero along the diagonal (i.e. for $r=q$). An example of the periapsis, $r_{\mathrm{p}}$, apoapsis, $r_{\mathrm{a}}$, and best parametrisation point, $q_\star$, are shown in red and blue. In this system of units, the Plummer characteristic radius, $\kappa$, lies at $r=1$.}
    \label{fig:epsB}
\end{figure}

Figure~\ref{fig:epsB} shows the behaviour of $\varepsilon\mathcal{B}_q(r)$ as a function of the parametrisation point $q$ and the star's radial location $r$. As the star moves between a periapsis and apoapsis, $r$ varies between $r_{\mathrm{p}}$ and $r_{\mathrm{a}}$, describing a horizontal segment in Fig.~\ref{fig:epsB}. Each parametrisation point $q$ defines a given horizontal segment, and the collection of these segments form a square (see the example in red). For a given parameter $q$, we define the `perturbation index' $P_q$ as the maximum value of $\varepsilon\mathcal{B}_q(r)$ for $r\in[r_{\mathrm{p}},r_{\mathrm{a}}]$:
\begin{equation}
    P_q = \max\Big[\varepsilon\mathcal{B}_q(r)\Big]\,,\quad r\in[r_{\mathrm{p}},r_{\mathrm{a}}] \,.
\end{equation}
There exists a unique point $r_{\mathrm{c}} > 0$ such that $\partial_r\varepsilon\mathcal{B}_q(r_{\mathrm{c}}) = 0$ and $\partial^2_{rr}\varepsilon\mathcal{B}_q(r_{\mathrm{c}}) < 0$, which corresponds to a local maximum. It can be shown than $r_{\mathrm{c}} > q$ (see Appendix~\ref{an:MaxExist}). If we restrict $r$ to lie in $[r_{\mathrm{p}},r_{\mathrm{a}}]$, then the maximum of $\varepsilon\mathcal{B}_q(r)$ can be reached at $r_{\mathrm{p}}$, $r_{\mathrm{a}}$, or $r_{\mathrm{c}}$, that is,
\begin{equation}
   P_q = 
   \left\{
   \begin{aligned}
       &\max\Big\{\varepsilon\mathcal{B}_q(r_{\mathrm{p}}),\varepsilon\mathcal{B}_q(r_{\mathrm{c}})\Big\}\text{ when }r_{\mathrm{c}} < r_{\mathrm{a}}\,,\\
       &\max\Big\{\varepsilon\mathcal{B}_q(r_{\mathrm{p}}),\varepsilon\mathcal{B}_q(r_{\mathrm{a}})\Big\}\text{ otherwise}\,.
   \end{aligned}
   \right.
\end{equation}
In particular, for $q=r_\mathrm{a}$, we always have $P_{\mathrm{a}} = \varepsilon\mathcal{B}_{\mathrm{a}}(r_{\mathrm{p}})$. The best parametrisation point $q=q_\star$ is that for which $P_q$ is minimum; this minimum value,  $P_\star$, is
\begin{equation}
    P_\star = \min\Big[P_q\Big],\quad q\in[r_{\mathrm{p}},r_{\mathrm{a}}]\,.
\end{equation}
By definition, $\varepsilon\mathcal{B}_{\mathrm{p}}(r_{\mathrm{p}})=0$, and $\varepsilon\mathcal{B}_q(r_{\mathrm{p}})$ is an increasing function of $q$. In contrast, $\varepsilon\mathcal{B}_q(r_{\mathrm{c}})$ and $\varepsilon\mathcal{B}_q(r_{\mathrm{a}})$ are decreasing function of $q$, and by definition $\varepsilon\mathcal{B}_{\mathrm{a}}(r_{\mathrm{a}})=0$ (see Appendix~\ref{an:Des_epsB>0}). Hence, the best parametrisation point $q_\star$ is
\begin{equation}
    \label{eq:qSTAR}
    q_\star = \Big\{q\text{ such that } \varepsilon\mathcal{B}_q(r_{\mathrm{p}}) = \varepsilon\mathcal{B}_q\big(\min(r_{\mathrm{c}},r_{\mathrm{a}})\big)\Big\}\,.
\end{equation}
We can compute it at an arbitrary precision using bisection.

\begin{figure}
    \includegraphics[width=\columnwidth]{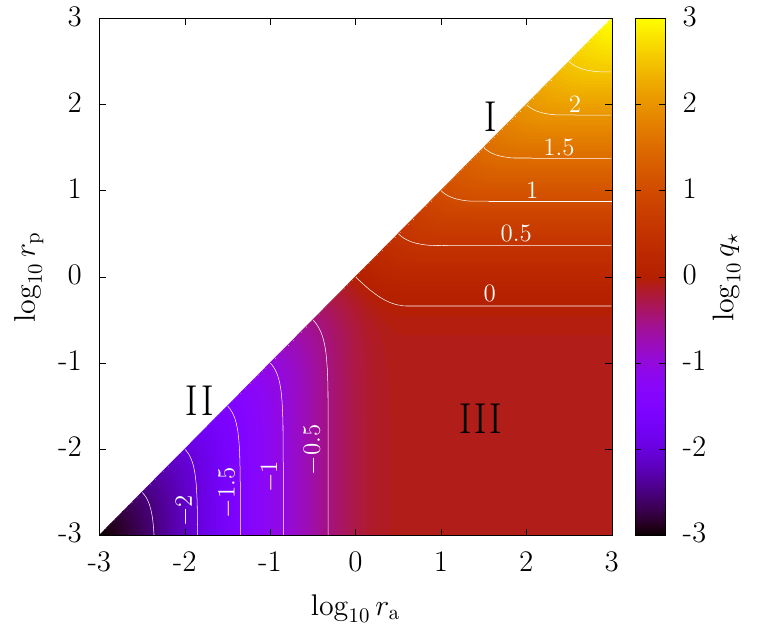}
    \caption{Optimal parametrisation point, $q_\star$, as a function of the periapsis and apoapsis of the orbit. Some level curves appear in white. Regions~I,~II, and~III are described in the text.}
    \label{fig:qstar}
\end{figure}

Figure~\ref{fig:qstar} shows the positions of the optimal parametrisation point $q_\star$ for orbits with periapsis and apoapsis distances ranging from $10^{-3}$ to $10^{3}$ (in our system of units such that the Plummer scale radius is $\kappa=1$). Three distinct regions can be identified: I.~for $r_{\mathrm{p}}>\kappa$, we have $q_\star\approx r_{\mathrm{p}}$; II.~for $r_{\mathrm{a}}<\kappa$, we have $q_\star\approx r_{\mathrm{a}}$; III.~for $r_{\mathrm{p}}<\kappa$ and $r_{\mathrm{a}}>\kappa$, we have $q_\star\approx\kappa$.

The performances of our splitting scheme can be evaluated by the size of the perturbation part $\varepsilon\mathcal{B}$ (see Sect.~\ref{sec:sympinteg}). Here, the total Hamiltonian function is $\mathcal{H}=\|\dot{\mathbf{r}}\|^2/2 + \Psi(\mathbf{r})$, where $\Psi(\mathbf{r})$ is the Plummer potential. If we consider a bound orbit with pericentre $r_{\mathrm{p}}$ and apocentre $r_{\mathrm{a}}$, then the constant value of $\mathcal{H}$ is automatically set to $\mathcal{H}=k$, where
\begin{equation}
    k = \frac{r_{\mathrm{p}}^2\sqrt{1+r_{\mathrm{a}}^2} - r_{\mathrm{a}}^2\sqrt{1+r_{\mathrm{p}}^2}}{(r_{\mathrm{a}}^2-r_{\mathrm{p}}^2)\sqrt{(1+r_{\mathrm{p}}^2)(1+r_{\mathrm{a}}^2)}} \in [-1,0)\,.
\end{equation}
For given values of $r_{\mathrm{p}}$ and $r_{\mathrm{a}}$, and a given parametrisation radius $q\in[r_{\mathrm{p}},r_{\mathrm{a}}]$, we define two quantities that characterise the performances of our splitting scheme: the size of the perturbation,
\begin{equation}
    \label{eq:Eq}
    E_q = -\frac{P_q}{k} > 0\,,
\end{equation}
and the relative magnitude of the perturbation,
\begin{equation}
    \label{eq:epsq}
    \varepsilon_q = \frac{P_q}{|k-P_q|} >0\,.
\end{equation}
Namely, we have $E_q\approx |\varepsilon\mathcal{B}/\mathcal{H}|$, and $\varepsilon_q \approx |\varepsilon\mathcal{B}/\mathcal{A}|$. If $E_q$ is small, then $E_q \approx \varepsilon_q$, and both Eqs.~\eqref{eq:Eq} and~\eqref{eq:epsq} quantify the relative magnitude of the perturbation. Instead, when $E_q$ is large, then $\varepsilon_q\approx 1$ (i.e. there is no hierarchy between the `perturbation' $\varepsilon\mathcal{B}$ and the `unperturbed part' $\mathcal{A}$ in the Hamiltonian function). In that case $E_q$ quantifies the magnitude of quantities that our splitting scheme artificially inserts in the Hamiltonian function, with no benefit on its performances.

\begin{figure}
    \includegraphics[width=\columnwidth]{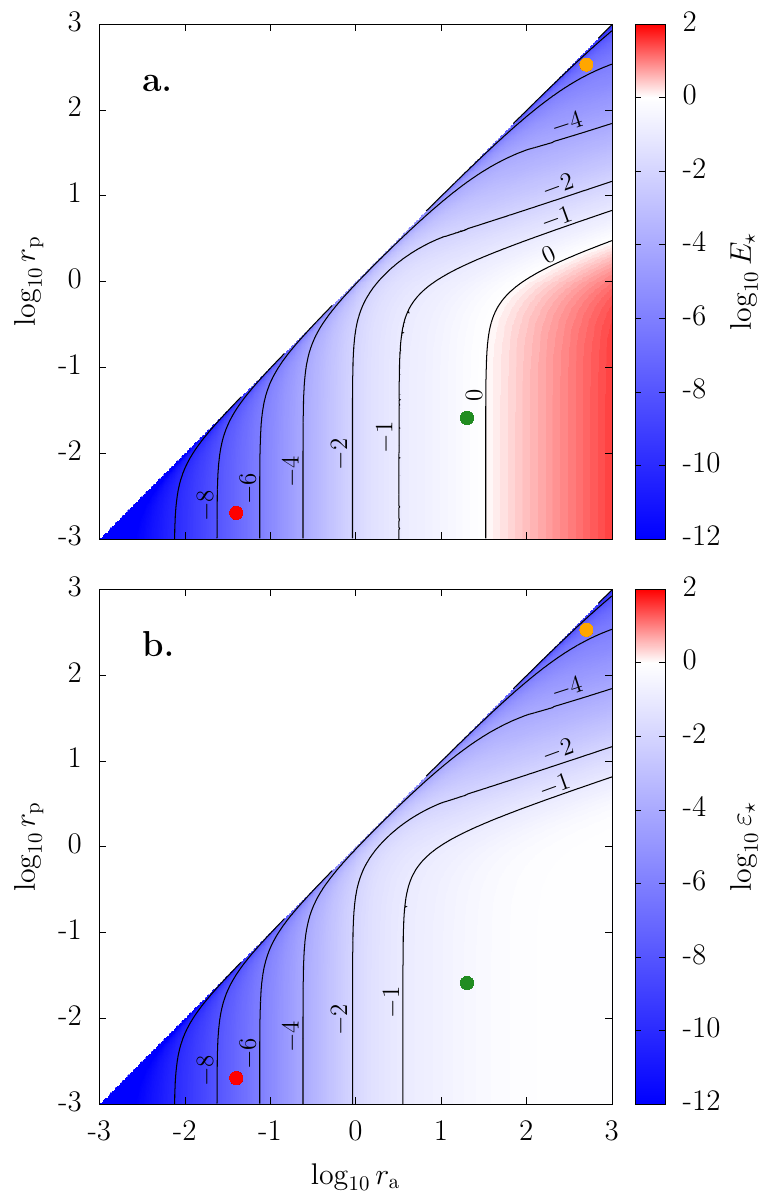}
    \caption{Size, $E_\star$, of the perturbation (panel~a) and relative magnitude, $\varepsilon_\star$, of the perturbation (panel~b) computed for the optimal parametrisation point, $q_\star$. Some level curves appear in black. The three coloured points were used for the numerical experiments described in Sect.~\ref{sec:perf} (see Figs.~\ref{fig:compInt_SABA1} and~\ref{fig:compInt_SABA3}).}
    \label{fig:epsilonstar}
\end{figure}

\begin{figure}
    \includegraphics[width=\columnwidth]{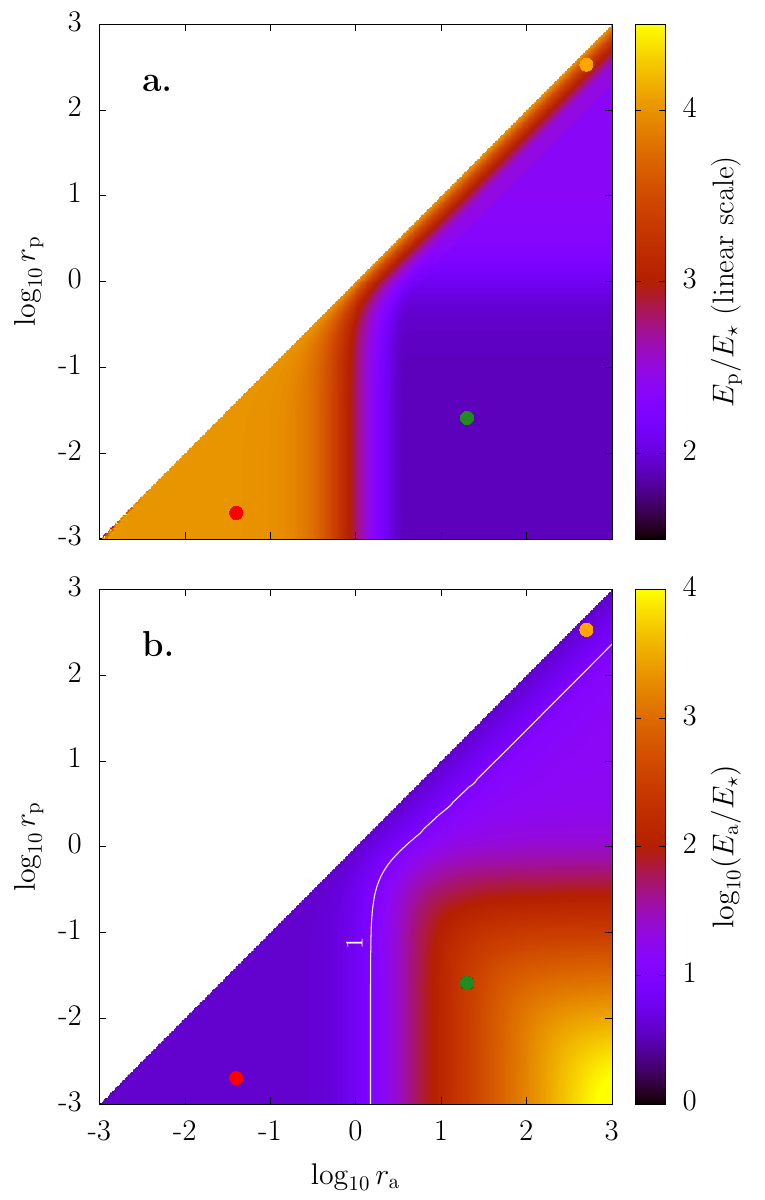}
    \caption{Size, $E,$ of the perturbation computed for $q=r_{\mathrm{p}}$ (panel~a) and $q=r_{\mathrm{a}}$ (panel~b) compared to the best parametrisation point, $q_\star$. The colour scale in panel~a is linear; the colour scale in panel~b is logarithmic. The level curve $E_{\mathrm{a}}/E_\star = 10$ is shown in white. The three coloured points were used for the numerical experiments described in Sect.~\ref{sec:perf} (see Figs.~\ref{fig:compInt_SABA1} and~\ref{fig:compInt_SABA3}).}
    \label{fig:epsilonratio}
\end{figure}

The $E_{\mathrm{p}}$, $E_{\mathrm{a}}$, $E_\star$, and $\varepsilon_{\mathrm{p}}$, $\varepsilon_{\mathrm{a}}$, $\varepsilon_\star$ are the coefficients obtained when using a parametrisation radius $q$ equal to $r_{\mathrm{p}}$, $r_{\mathrm{a}}$, or $q_\star$, respectively. Figure~\ref{fig:epsilonstar} shows the value of $E_\star$ and $\varepsilon_\star$ as a function of $r_{\mathrm{a}}$ and $r_{\mathrm{p}}$. As expected, $\varepsilon_\star$ is very small when $r_{\mathrm{p}}\gg \kappa$ (Keplerian regime) and when $r_{\mathrm{a}}\ll \kappa$ (harmonic regime). We also obtain good performances when $r_{\mathrm{a}}\approx\kappa$ and/or $r_{\mathrm{p}}\approx\kappa$, with values of $\varepsilon_\star$ of the order of $10^{-1}$ or $10^{-2}$. Yet, we expect no advantage of using our splitting scheme for highly eccentric orbits crossing the Plummer radius $\kappa$ (i.e. for $r_{\mathrm{a}}\gg\kappa$  and $r_{\mathrm{p}}<\kappa$; see the red region in Fig.~\ref{fig:epsilonstar}a), for which $E_\star>10$, and therefore $\varepsilon_\star\approx 1$.

This analysis shows that the best way to set up our splitting scheme depends on the properties of the trajectory followed by the star. However, in general we cannot know a priori what the star's trajectory will be --- we only know its initial conditions. Therefore, we must determine whether choosing the best parametrisation point $q_\star$ is critical to get the best performances of the integrator, or whether any other point along the star's orbit (e.g. $q=r_{\mathrm{p}}$ or $r_{\mathrm{a}}$, or the initial radial location) could give approximately similar performances. 

Figure~\ref{fig:epsilonratio} shows how the perturbation sizes $E_{\mathrm{p}}$ and $E_{\mathrm{a}}$ behave relative to $E_\star$ as a function of $r_{\mathrm{a}}$ and $r_{\mathrm{p}}$. Figure~\ref{fig:epsilonratio}a shows that $E_\star$ and $E_{\mathrm{p}}$ only differ by a factor close to unity. Namely, $E_{\mathrm{p}}/E_\star\approx 2.5$ in region~I, $E_{\mathrm{p}}/E_\star\approx 3.5$ in region~II, and $E_{\mathrm{p}}/E_\star\approx 2$ in region~III. Moreover, we note that the largest ratio occurs in region~II, where $E_\star$ is the smallest (see Fig.~\ref{fig:epsilonstar}b). We conclude that using the parametrisation point $q=r_{\mathrm{p}}$ instead of $q=q_\star$ would not damage too much the performances of our splitting scheme. In contrast, Fig.~\ref{fig:epsilonratio}b shows that $E_{\mathrm{a}}$ is orders of magnitudes larger that $E_\star$ almost everywhere, except in region~II and near the diagonal (for which $r_{\mathrm{a}}\approx r_{\mathrm{p}}$ anyway).

We conclude that choosing an adequate parametrisation point $q\in [r_{\mathrm{p}},r_{\mathrm{a}}]$ is indeed critical. Yet, choosing simply $q=r_{\mathrm{p}}$ always give relevant results; it is therefore an easy choice to optimise the performance of the symplectic integrator developed in Sect.~\ref{sec:IsoSymp}. We note that it remains true for unbound orbits.

\section{Performance for stellar dynamics}
\label{sec:perf}
In this section we illustrate the properties obtained in Sect.~\ref{sec:Plummer} using numerical integrations. In galactic dynamics, the Plummer potential is often used to describe stellar clusters, in particular globular clusters. To illustrate the performances of our splitting scheme, we apply the symplectic integrator presented in Sect.~\ref{sec:IsoSymp} to a star orbiting in NGC~4372, assuming a Plummer distribution for the cluster. The total mass of NGC~4372 is $M=1.9\times10^5~\mathcal{M}_\sun$, and its half-mass radius is $r_{1/2}=8.34~\text{pc}$, according to~\cite{Baumgardt_2018}~\footnote{Specifically, the adopted values were taken from the edition available at~\url{https://people.smp.uq.edu.au/HolgerBaumgardt/globular/parameter.html} as of September 30, 2024.}. The Plummer mass parameter, $\eta$, is related to the cluster's mass by $\eta = GM$, where $G\mathcal{M}_\sun=4.49850\times10^{-3}~\text{pc}^3~\text{Myr}^{-2}$ is the Solar gravitational parameter~\citep{Prsa_2016}. The Plummer scale radius parameter, $\kappa$, is related to the half-mass radius by $r_{1/2}\simeq1.305\kappa$~\citep{Heggie_2003}. We do not mean here to reproduce the actual orbital evolution of stars within the NGC~4372 globular cluster, but only to illustrate our results using physically motivated parameter values for $\eta$ and $\kappa$.

We considered three different orbits. These points are taken: in Region~I ($r_{\mathrm{p}}\gg\kappa$) and close to a circular orbit; in Region~II ($r_{\mathrm{a}}<\kappa$); and in Region~III ($r_{\mathrm{p}}<\kappa$ and $r_{\mathrm{a}}>\kappa$). We call them respectively Orbit~$\#1$, Orbit~$\#2$, and Orbit~$\#3$. Their radial periods are respectively $T\approx 30$~Gyr, $1.73$~Myr, and $112$~Myr. These trajectories are designed to test our integration scheme in a large variety of regimes, even extreme regimes (e.g. Orbit~$\#1$ may not be physically realistic).

\begin{figure*}
    \includegraphics[width=\textwidth]{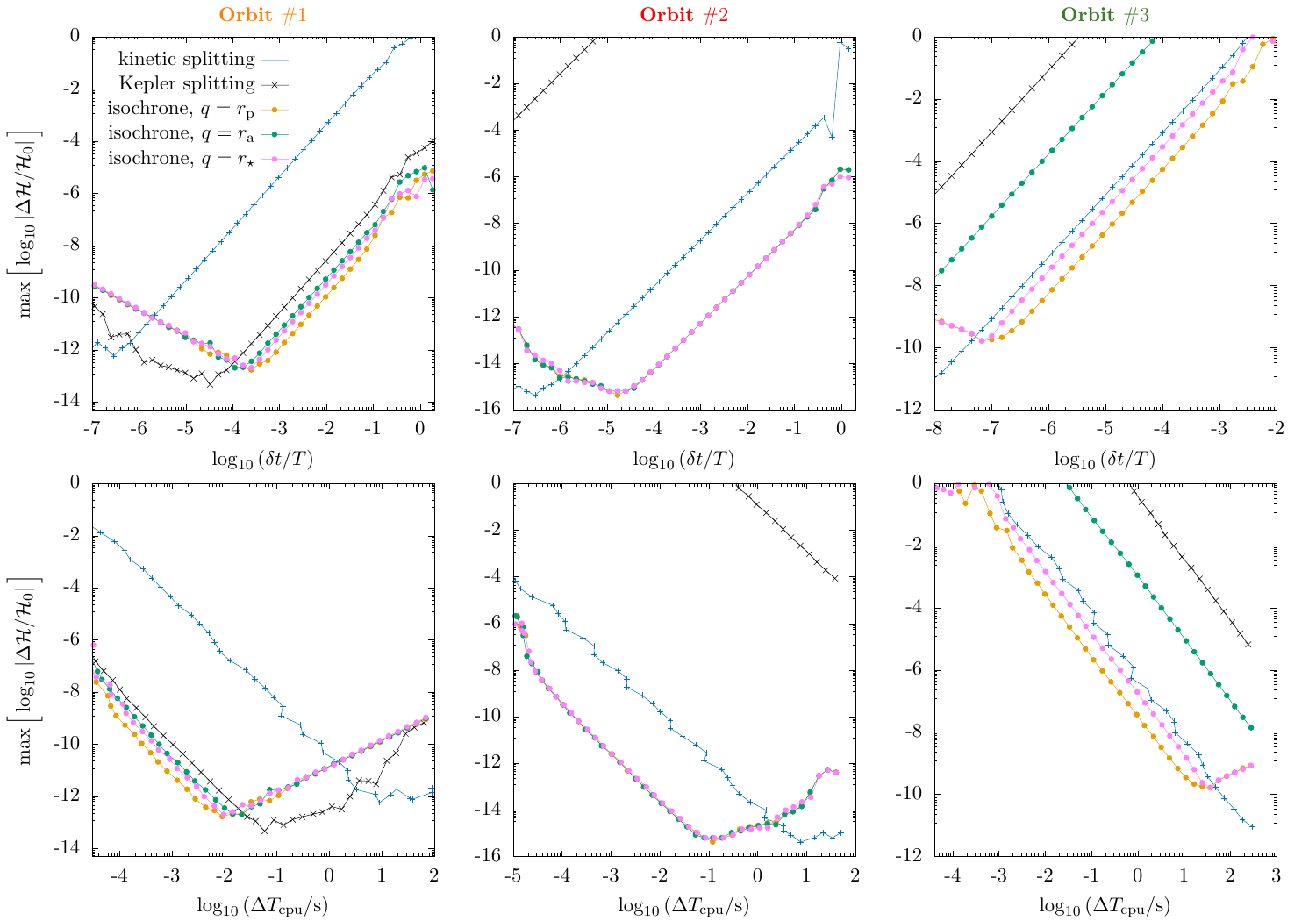}
    \caption{Maximum relative variation in the Hamiltonian as a function of the integration time step over the radial period (\textit{top row}) and as a function of the CPU integration time (\textit{bottom row}). Each column corresponds to a different orbit (orange, red, and green points in Figs.~\ref{fig:epsilonstar} and~\ref{fig:epsilonratio}), integrated over two radial periods with $\mathcal{SABA}_1$. The blue curve and the black curve show the results obtained when using kinetic splitting and Kepler splitting. The orange, green, and pink curves correspond to the isochrone splitting with parametrisation points $q=r_{\mathrm{p}}$, $q=r_{\mathrm{a}}$, and $q=r_\star$, respectively.}
    \label{fig:compInt_SABA1}
\end{figure*}

\begin{figure*}
    \includegraphics[width=\textwidth]{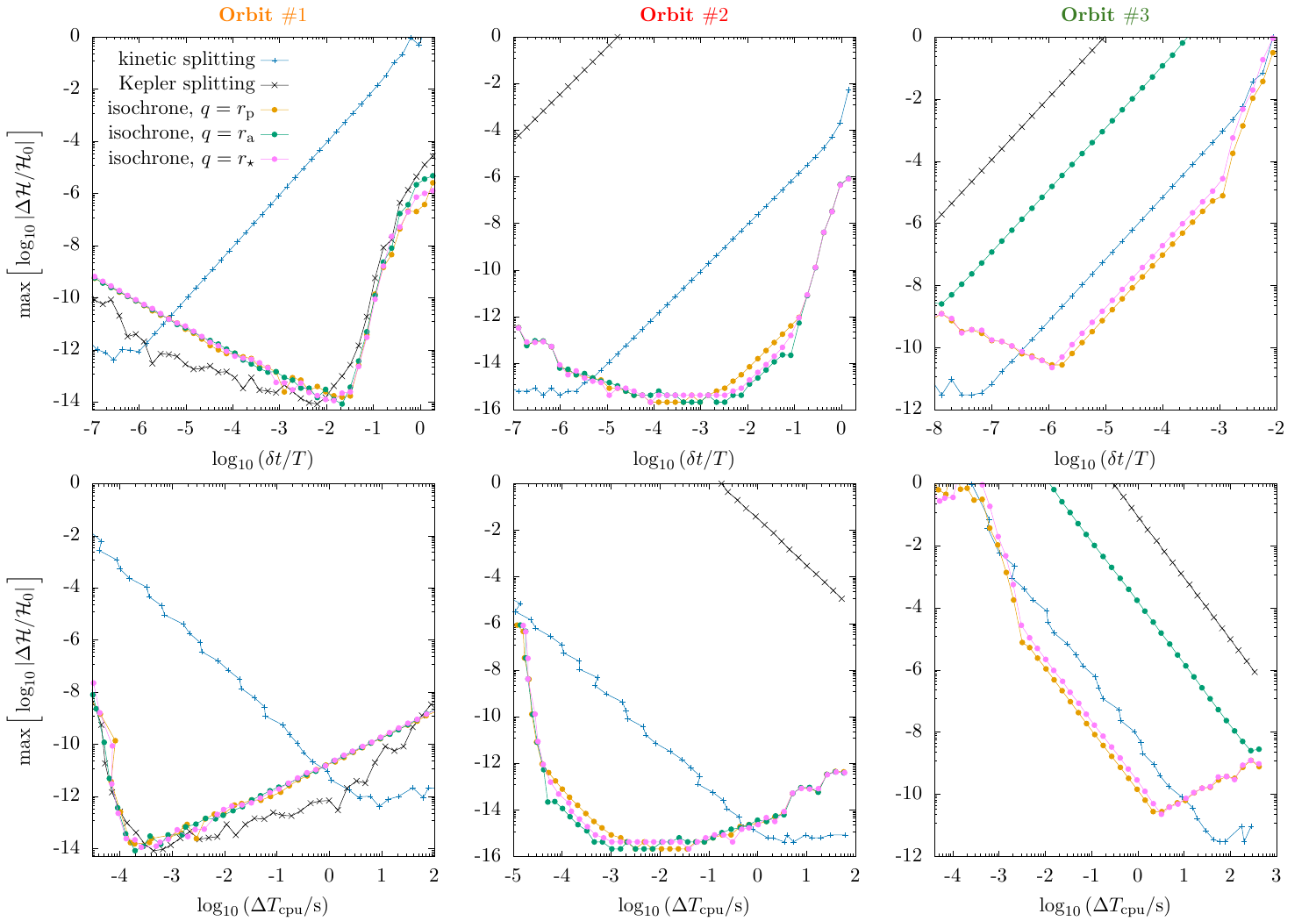}
    \caption{Same as Fig.~\ref{fig:compInt_SABA1} but using the $\mathcal{SABA}_{3}$ integrator.}
    \label{fig:compInt_SABA3}
\end{figure*}

Figure~\ref{fig:compInt_SABA1} shows the performances of the isochrone splitting scheme, as compared to the kinetic and Kepler splitting schemes. Integrations are performed with the widely used second-order symplectic integrator given in Eq.~\eqref{eq:Oleapfrog}. This integrator is called $\mathcal{SABA}_1$ by~\cite{Laskar_2001} and it has errors of order $\mathcal{O}(\varepsilon\delta t^2)$. Kinetic splitting corresponds to $\Phi = 0$ in Eqs.~\eqref{eq:A-HAM} and~\eqref{eq:eB-HAM}; when used with $\mathcal{SABA}_1$, it corresponds to the standard Leapfrog integrator (e.g. implemented in the Python package \texttt{galpy}). Kepler splitting uses $\Phi=\mu/r$, where we chose $\mu=GM$. Kepler splitting involves a state-of-the-art two-body propagator, with in particular the Gauss functions~$f$~and~$g$ (see e.g.~\citealp{Danby_1988}).

For Orbits~$\#1$ and~$\#2$, the choice of parametrisation point $q$ for the isochrone potential does not affect much the integrator's performance, as expected. The classic Leapfrog is outperformed by several orders of magnitude. Depending on the level of energy conservation requested, time step $100$ to $1000$ times larger can be used, with a noticeable benefit on the computation time (say, a factor of $10$ to $100$; see the bottom panels). Indeed, for these two orbits, the magnitude $\varepsilon$ of the perturbation when using the isochrone splitting scheme is very small (see Sect.~\ref{sec:Plummer}), which allows us to take full advantage of the hierarchical symplectic integrator.

For Orbit~$\#3$, the choice of parametrisation point $q$ is important: if we choose $q=r_{\mathrm{a}}$, then the isochrone splitting scheme performs poorly --- it is even less efficient than kinetic splitting. If we choose $q=r_{\mathrm{p}}$ or $q=r_\star$ instead, we obtain a slightly better conservation of energy than kinetic splitting (top right panel), but an equivalent cost in terms of CPU time (bottom right panel). Indeed, for Orbit~$\#3$, the perturbation magnitude $\varepsilon$ is close to unity (see the green point in Figs.~\ref{fig:epsilonstar} and~\ref{fig:epsilonratio}). In such a case, there is no advantage in using a symplectic splitting scheme based on a particular hierarchy between the two terms of the Hamiltonian function.

Surprisingly, we note that for Orbit~$\#3$ the isochrone splitting scheme for $q=r_{\mathrm{p}}$ performs slightly better than for $q=q_\star$; this is likely due to particular terms in the remainders of the integrator (see e.g.~\citealp{Laskar_2001}), which lose their well-behaved hierarchy when $\varepsilon\approx 1$. This property confirms our conclusion from Sect.~\ref{sec:Plummer} that setting $q=r_\mathrm{p}$ is an simple choice that optimises the isochrone splitting scheme.

As expected, Kepler splitting performs similarly as the isochrone splitting for Orbit~$\#1$ (see the black curve in Fig.~\ref{fig:compInt_SABA1}), because this orbit is far from the centre of the cluster. However, Kepler splitting shows very poor performances for Orbits~$\#2$ and~$\#3$, which are strongly non-Keplerian. This illustrates that the isochrone splitting is a generalisation of Kepler splitting: it works equally well for perturbed Keplerian orbits but it has a wider range of applications.

The improved hierarchy (i.e. small value of $\varepsilon$) when using the isochrone splitting implies that if a better conservation of energy is needed, more advanced integrators will converge faster to a high precision. This property is illustrated in Fig.~\ref{fig:compInt_SABA3}, showing the results obtained with the integrator $\mathcal{SABA}_3$, which has errors of the form $\mathcal{O}(\varepsilon\delta t^6) + \mathcal{O}(\varepsilon^2\delta t^2)$. For large time steps, the slope~$6$ is clearly visible (as compared to the slope~$2$ in Fig.~\ref{fig:compInt_SABA1}). In contrast, the slope remains unchanged for kinetic splitting, because the second error terms in $\mathcal{O}(\varepsilon^2\delta t^2)$ dominates. However, as discussed in Sect.~\ref{sec:intro} high precision integrations are generally not needed in the context of galactic dynamics.

\section{Discussion}
\label{sec:discussion}
Similarly to other splitting schemes (see e.g.~\citealp{Hernandez_2017}), the choice of the isochrone parameters $\mu$ and $b$ made in Eq.~\eqref{eq:mub-generalcase} is not unique. We used Eq.~\eqref{eq:mub-generalcase} as an attempt to minimise the perturbation Hamiltonian $\varepsilon\mathcal{B}$ in a simple way, but other choices for $\mu$ and $b$ could possibly yield equal performances, or even better performances than what we obtained in Sect.~\ref{sec:Plummer}.

\begin{figure*}
     \includegraphics[width=\textwidth]{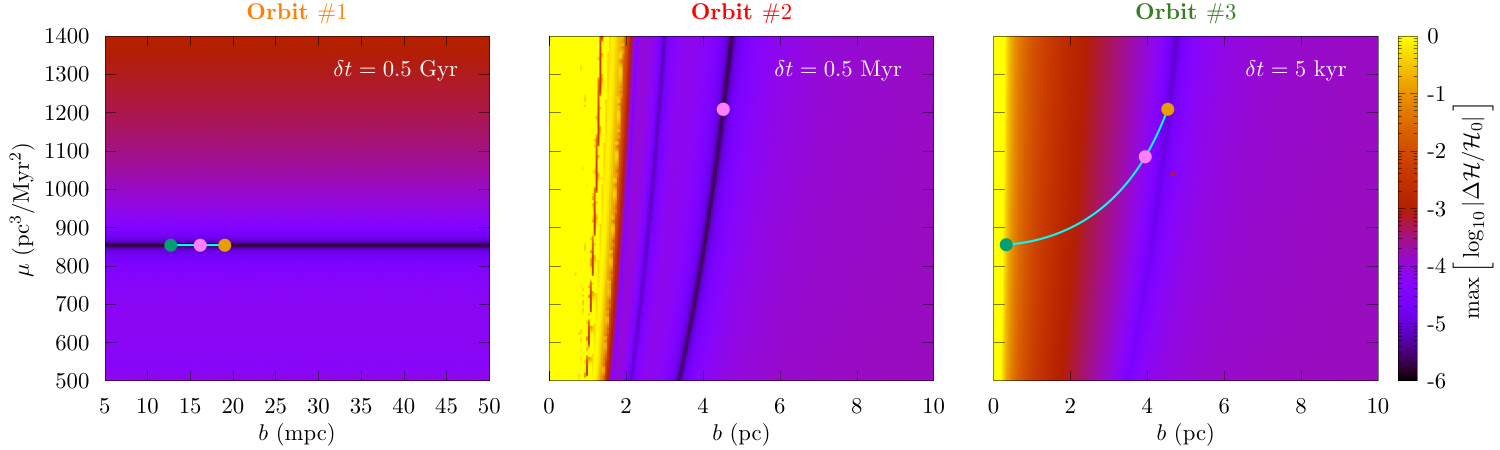}
     \caption{Maximum relative variation in the Hamiltonian as a function of the parameters $\mu$ and $b$ of the isochrone potential used for the splitting. Each column corresponds to a different orbit (orange, red, and green points in Figs.~\ref{fig:epsilonstar} and~\ref{fig:epsilonratio}), integrated with $\mathcal{SABA}_{1}$ over two radial periods for a given integration time step, $\delta t$. The orange, green, and pink points correspond to the isochrone splitting with parametrisation points $q=r_{\mathrm{p}}$, $q=r_{\mathrm{a}}$, and $q=r_{\star}$, respectively (the points overlap for Orbit~$\#2$), while the cyan curve shows all possible parametrisation points, $q$, between the periapsis distance and apoapsis distance, following Eq.~\eqref{eq:mub}.}
    \label{fig:mub-alternative_SABA1}
\end{figure*}

Figure~\ref{fig:mub-alternative_SABA1} illustrates the integrator's performance as a function of the splitting parameters, which do not specifically follow Eq.~\eqref{eq:mub-generalcase}, for the three orbits presented in Sect.~\ref{sec:perf}. For all three orbits, we observe that there is a region where the performance of the integrator is much better. The choice made in Sect.~\ref{sec:IsoSymp} falls within this region for a well-chosen parametrisation point $q$. As we mentioned in Sect.~\ref{sec:perf}, the parametrisation at the periapsis distance slightly outperforms the parametrisation at the optimal point defined in Eq.~\eqref{eq:qSTAR} for the Orbit~$\#3$. Indeed, the characteristic size of the perturbation for Orbit~$\#3$ is $\varepsilon\approx 1$ (see Fig.~\ref{fig:epsilonstar}), so we expect a poor convergence of the remainders as expressed in powers of $\varepsilon$. As a result the exact minimum of $\varepsilon$ does not necessarily gives exactly the best performances of the integrator, depending on the splitting scheme used, but it comes close.

\section{Summary and conclusions}\label{sec:conclusion}

In many studies of stellar and galactic dynamics, it is necessary to numerically integrate a large number of test particles (typically stars) in some smoothed gravitational potential. Symplectic integrators are widely used in this context because of their exquisite stability properties. To date, the vast majority of studies have used the kinetic symplectic splitting scheme, for which the integrator alternates between a linear drift and a velocity kick (see e.g.~\citealp{MartinezBarbosa-Brown-PortegiesZwart_2015,Pouliasis-DiMatteo-Haywood_2017,Ferrone-etal_2023}). However, much more efficient symplectic integrators have been developed for cases where the Hamiltonian has a particular hierarchical structure (e.g. a dominant part plus a perturbation; see~\citealp{Wisdom_1991,McLachlan_1995,Laskar_2001}). We investigated whether such integrators can be advantageous for stellar dynamics with an appropriate choice of splitting.

In this regard, Hénon's isochrone potential seems very promising because it shares many properties with complex galactic potentials (e.g. it is harmonic at short distances and Keplerian at large distances), and the isochrone dynamics is fully integrable analytically. Therefore, we used Hénon's isochrone potential as a basis for a new splitting scheme, for which the integrator alternates between an isochrone drift (Hamiltonian $\mathcal{A}$) and a velocity kick (Hamiltonian $\varepsilon\mathcal{B}$). To this end, we have expressed all solutions of the isochrone dynamics --- including unbound trajectories --- in an explicit form, by introducing an `eccentric anomaly' similar to the Keplerian one. The problem then becomes one of selecting the mass parameter, $\mu$, and the characteristic distance, $b,$ of the isochrone potential used in the integration scheme such that the characteristic magnitude, $\varepsilon,$ of the perturbations is minimised.

As a first application of this splitting scheme, we integrated the motion of stars in a Plummer potential --- which is an essential component of most galactic potentials. In this case, we have shown that a judicious choice for $\mu$ and $b$ is to set $\varepsilon\mathcal{B}$ and its first derivative to zero at the star's periapsis. This way, $\varepsilon$ takes very small values for a large fraction of trajectories: typically $10^{-6}$ for stars near the centre of the mass distribution (harmonic regime) and $10^{-3}$ for stars in the periphery of the distribution (Keplerian regime). Poorer performances are obtained for stars on very elongated orbits that cross the Plummer characteristic radius, for which $\varepsilon\sim 1$.

Numerical experiments show that, depending on the dynamical regime of the star considered, this new splitting scheme can outperform the standard Leapfrog integrator by several orders of magnitude. As a result, computations are faster by a factor of $10$ to $100$ for the same energy conservation. In the worst cases --- very elongated orbits that cross both inner and outer regions of the Plummer distribution --- performances are equivalent to those obtained with previous integration methods.

The isochrone splitting scheme introduced here remains to be explored for more general
potentials among those used in galactic dynamics. In these future applications, the parameters $\mu$ and $b$ will need to be chosen in an adequate way, depending on the potentials included. The results obtained here give us confidence that good performances will be obtained in future applications. We expect clear improvements compared to previous methods at least for some classes of stellar trajectories.

According to the specific implementation methodology, the method presented here could be improved in many ways in the future. For instance, analogous versions of the Gauss functions $f$ and $g$ could be derived for the isochrone dynamics, and round-off errors and computation times could be optimised more than in our proof-of-concept implementation presented here. We leave such refinements for future studies.

\begin{acknowledgements}
   We thank Alain Vienne for useful discussions about numerical cancellation errors. We also thank the anonymous reviewer for his/her very thorough review that greatly improved the clarity of our manuscript. This work was supported by the Programme de Planétologie (PNP) of CNRS/INSU, co-funded by CNES.
\end{acknowledgements}

\bibliographystyle{aa}
\bibliography{isochrone}

\begin{thebibliography}{58}
\expandafter\ifx\csname natexlab\endcsname\relax\def\natexlab#1{#1}\fi

\bibitem[{{Allen} \& {Santillan}(1991)}]{Allen-Santillan_1991}
{Allen}, C. \& {Santillan}, A. 1991, \rmxaa, 22, 255

\bibitem[{{Baba} {et~al.}(2024){Baba}, {Tsujimoto}, \&
  {Saitoh}}]{Baba-Tsujimoto-Saitoh_2024}
{Baba}, J., {Tsujimoto}, T., \& {Saitoh}, T.~R. 2024, The Astrophysical Journal
  Letters, 976, L29

\bibitem[{{Baumgardt} \& {Hilker}(2018)}]{Baumgardt_2018}
{Baumgardt}, H. \& {Hilker}, M. 2018, \mnras, 478, 1520

\bibitem[{{Bertrand}(1873)}]{Bertrand_1873}
{Bertrand}, J.~L.~F. 1873, Comptes rendus hebdomadaires des séances de
  l'Acad{\'e}mie des Sciences, 77, 849

\bibitem[{{Binney}(2016)}]{Binney_2014}
{Binney}, J. 2016, in Une vie dédiée aux systèmes dynamiques : Hommage {\`a}
  Michel H{\'e}non, ed. J.-M. {Alimi}, R.~{Mohayaee}, \& J.~{Perez} (Hermann),
  99--109

\bibitem[{{Binney} \& {Tremaine}(2008)}]{Binney-Scott_2008}
{Binney}, J. \& {Tremaine}, S. 2008, {Galactic Dynamics: Second Edition}
  (Princeton University Press)

\bibitem[{Blanes {et~al.}(2013)Blanes, Casas, Farrés, Laskar, Makazaga, \&
  Murua}]{Blanes-etal_2013}
Blanes, S., Casas, F., Farrés, A., {et~al.} 2013, Applied Numerical
  Mathematics, 68, 58

\bibitem[{{Boulila} {et~al.}(2018){Boulila}, {Laskar}, {Haq}, {Galbrun}, \&
  {Hara}}]{Boulila_2018}
{Boulila}, S., {Laskar}, J., {Haq}, B.~U., {Galbrun}, B., \& {Hara}, N. 2018,
  Global and Planetary Change, 165, 128

\bibitem[{{Bovy}(2015)}]{Bovy_2015}
{Bovy}, J. 2015, \apjs, 216, 29

\bibitem[{{Carit{\'a}} {et~al.}(2018){Carit{\'a}}, {Rodrigues}, {Puerari}, \&
  {Schiavo}}]{Carita-Rodrigues-Puerari-Schiavo_2018}
{Carit{\'a}}, L.~A., {Rodrigues}, I., {Puerari}, I., \& {Schiavo}, L. E. C.~A.
  2018, \na, 60, 48

\bibitem[{{Chambers}(1999)}]{Chambers_1999}
{Chambers}, J.~E. 1999, \mnras, 304, 793

\bibitem[{{Danby}(1988)}]{Danby_1988}
{Danby}, J. M.~A. 1988, {Fundamentals of celestial mechanics} (Atlantic Books)

\bibitem[{{Dinescu} {et~al.}(1999){Dinescu}, {Girard}, \& {van
  Altena}}]{Dinescu-Girard-vanAltena_1999}
{Dinescu}, D.~I., {Girard}, T.~M., \& {van Altena}, W.~F. 1999, \aj, 117, 1792

\bibitem[{{Farr{\'e}s} {et~al.}(2013){Farr{\'e}s}, {Laskar}, {Blanes}, {Casas},
  {Makazaga}, \& {Murua}}]{Farres-etal_2013}
{Farr{\'e}s}, A., {Laskar}, J., {Blanes}, S., {et~al.} 2013, Celestial
  Mechanics and Dynamical Astronomy, 116, 141

\bibitem[{{Ferrone} {et~al.}(2023){Ferrone}, {Di Matteo},
  {Mastrobuono-Battisti}, {Haywood}, {Snaith}, {Montuori}, {Khoperskov}, \&
  {Valls-Gabaud}}]{Ferrone-etal_2023}
{Ferrone}, S., {Di Matteo}, P., {Mastrobuono-Battisti}, A., {et~al.} 2023,
  \aap, 673, A44

\bibitem[{{Forest} \& {Ruth}(1990)}]{Forest_1990}
{Forest}, E. \& {Ruth}, R.~D. 1990, Physica D Nonlinear Phenomena, 43, 105

\bibitem[{{Gardner} {et~al.}(2011){Gardner}, {Nurmi}, {Flynn}, \&
  {Mikkola}}]{Gardner-Nurm-Flynn-Mikkola_2011}
{Gardner}, E., {Nurmi}, P., {Flynn}, C., \& {Mikkola}, S. 2011, Monthly Notices
  of the Royal Astronomical Society, 411, 947

\bibitem[{{Greiner}(1987)}]{Greiner_1987}
{Greiner}, J. 1987, Celestial mechanics, 40, 171

\bibitem[{{Hairer} {et~al.}(2006){Hairer}, {Wanner}, \&
  {Lubich}}]{Hairer-Wanner-Lubich_2006}
{Hairer}, E., {Wanner}, G., \& {Lubich}, C. 2006, Geometric Numerical
  Integration (Springer Berlin, Heidelberg)

\bibitem[{{Heggie} \& {Hut}(2003)}]{Heggie_2003}
{Heggie}, D. \& {Hut}, P. 2003, {The Gravitational Million-Body Problem: A
  Multidisciplinary Approach to Star Cluster Dynamics} (Cambridge University
  Press)

\bibitem[{{H{\'e}non}(1959{\natexlab{a}})}]{Henon_1959_2}
{H{\'e}non}, M. 1959{\natexlab{a}}, Annales d'Astrophysique, 22, 491

\bibitem[{{H{\'e}non}(1959{\natexlab{b}})}]{Henon_1959_1}
{H{\'e}non}, M. 1959{\natexlab{b}}, Annales d'Astrophysique, 22, 126

\bibitem[{{Hernandez} \& {Dehnen}(2017)}]{Hernandez_2017}
{Hernandez}, D.~M. \& {Dehnen}, W. 2017, \mnras, 468, 2614

\bibitem[{{Hernandez} \& {Dehnen}(2024)}]{Hernandez-Dehnen_2024}
{Hernandez}, D.~M. \& {Dehnen}, W. 2024, \mnras, 530, 3870

\bibitem[{{Hernandez} {et~al.}(2020){Hernandez}, {Hadden}, \&
  {Makino}}]{Hernandez-etal_2020}
{Hernandez}, D.~M., {Hadden}, S., \& {Makino}, J. 2020, \mnras, 493, 1913

\bibitem[{{Ibata} {et~al.}(2019){Ibata}, {Malhan}, \&
  {Martin}}]{Ibata-Malhan-Martin_2019}
{Ibata}, R.~A., {Malhan}, K., \& {Martin}, N.~F. 2019, \apj, 872, 152

\bibitem[{{Kaib} {et~al.}(2011){Kaib}, {Ro\v{s}kar}, \&
  {Quinn}}]{Kaib-Roskar-Quinn_2011}
{Kaib}, N.~A., {Ro\v{s}kar}, R., \& {Quinn}, T. 2011, Icarus, 215, 491

\bibitem[{{Khalil} {et~al.}(2024){Khalil}, {Famaey}, {Monari}, {Bernet},
  {Siebert}, {Ibata}, {Thomas}, {Ramos}, {Antoja}, {Li}, {Rozier}, \&
  {Romero-G{\'o}mez}}]{Khalil_2024}
{Khalil}, Y.~R., {Famaey}, B., {Monari}, G., {et~al.} 2024, A non-axisymmetric
  potential for the Milky Way disk

\bibitem[{{Khoperskov} {et~al.}(2020){Khoperskov}, {Di Matteo}, {Haywood},
  {G{\'o}mez}, \& {Snaith}}]{Khoperskov-DiMatteo-Haywood-Gomez-Snaith_2020}
{Khoperskov}, S., {Di Matteo}, P., {Haywood}, M., {G{\'o}mez}, A., \& {Snaith},
  O.~N. 2020, \aap, 638, A144

\bibitem[{{Kinoshita} {et~al.}(1990){Kinoshita}, {Yoshida}, \&
  {Nakai}}]{Kinoshita-Yoshida-Nakai_1990}
{Kinoshita}, H., {Yoshida}, H., \& {Nakai}, H. 1990, Celestial Mechanics and
  Dynamical Astronomy, 50, 59

\bibitem[{{Laskar} \& {Robutel}(2001)}]{Laskar_2001}
{Laskar}, J. \& {Robutel}, P. 2001, Celestial Mechanics and Dynamical
  Astronomy, 80, 39

\bibitem[{{Mart{\'\i}nez-Barbosa} {et~al.}(2015){Mart{\'\i}nez-Barbosa},
  {Brown}, \& {Portegies Zwart}}]{MartinezBarbosa-Brown-PortegiesZwart_2015}
{Mart{\'\i}nez-Barbosa}, C.~A., {Brown}, A.~G.~A., \& {Portegies Zwart}, S.
  2015, \mnras, 446, 823

\bibitem[{{Mart{\'i}nez-Barbosa} {et~al.}(2017){Mart{\'i}nez-Barbosa},
  {J{\'i}lkov{\'a}}, {Portegies Zwart}, \&
  {Brown}}]{MartinezBarbosa-Jilkov-PortegiesZwart-Brown_2017}
{Mart{\'i}nez-Barbosa}, C.~A., {J{\'i}lkov{\'a}}, L., {Portegies Zwart}, S., \&
  {Brown}, A. G.~A. 2017, Monthly Notices of the Royal Astronomical Society,
  464, 2290

\bibitem[{{McLachlan}(1995)}]{McLachlan_1995}
{McLachlan}, R.~I. 1995, BIT numerical mathematics, 35, 258–268

\bibitem[{{Miyamoto} \& {Nagai}(1975)}]{Miyamoto-Nagai_1975}
{Miyamoto}, M. \& {Nagai}, R. 1975, Astronomical Society of Japan, 27, 533

\bibitem[{{Paczynski}(1990)}]{Paczynski_1990}
{Paczynski}, B. 1990, Astrophysical Journal, 348, 485

\bibitem[{{Pascale} {et~al.}(2022){Pascale}, {Nipoti}, \&
  {Ciotti}}]{Pascale-Nipoti-Ciotti_2022}
{Pascale}, R., {Nipoti}, C., \& {Ciotti}, L. 2022, \mnras, 509, 1465

\bibitem[{{P{\'e}rez-Villegas} {et~al.}(2018){P{\'e}rez-Villegas}, {Rossi},
  {Ortolani}, {Casotto}, {Barbuy}, \&
  {Bica}}]{PerezVillegas-Rossi-Ortolani-Casotto-Barbuy-Bica_2018}
{P{\'e}rez-Villegas}, A., {Rossi}, L., {Ortolani}, S., {et~al.} 2018, \pasa,
  35, e021

\bibitem[{{Pichardo} {et~al.}(2004){Pichardo}, {Martos}, \&
  {Moreno}}]{Pichardo-Martos-Moreno_2004}
{Pichardo}, B., {Martos}, M., \& {Moreno}, E. 2004, \apj, 609, 144

\bibitem[{{Plummer}(1911)}]{Plummer_1911}
{Plummer}, H.~C. 1911, MNRAS, 71, 460

\bibitem[{{Portegies Zwart} \& {Boekholt}(2014)}]{PortegiesZwart-Boekholt_2014}
{Portegies Zwart}, S. \& {Boekholt}, T. 2014, \apjl, 785, L3

\bibitem[{{Pouliasis} {et~al.}(2017){Pouliasis}, {Di Matteo}, \&
  {Haywood}}]{Pouliasis-DiMatteo-Haywood_2017}
{Pouliasis}, E., {Di Matteo}, P., \& {Haywood}, M. 2017, Astronomy \&
  Astrophysics, 598, A66

\bibitem[{{Pr{\v{s}}a} {et~al.}(2016){Pr{\v{s}}a}, {Harmanec}, {Torres},
  {Mamajek}, {Asplund}, {Capitaine}, {Christensen-Dalsgaard}, {Depagne},
  {Haberreiter}, {Hekker}, {Hilton}, {Kopp}, {Kostov}, {Kurtz}, {Laskar},
  {Mason}, {Milone}, {Montgomery}, {Richards}, {Schmutz}, {Schou}, \&
  {Stewart}}]{Prsa_2016}
{Pr{\v{s}}a}, A., {Harmanec}, P., {Torres}, G., {et~al.} 2016, \aj, 152, 41

\bibitem[{{Ramond} \& {Perez}(2020)}]{Ramond-Perez_2020}
{Ramond}, P. \& {Perez}, J. 2020, Celestial Mechanics and Dynamical Astronomy,
  132, 22

\bibitem[{{Ramond} \& {Perez}(2021)}]{Ramond-Perez_2021}
{Ramond}, P. \& {Perez}, J. 2021, Journal of Mathematical Physics, 62, 112704

\bibitem[{{Rein} \& {Tamayo}(2015)}]{Rein-Tamayo_2015}
{Rein}, H. \& {Tamayo}, D. 2015, \mnras, 452, 376

\bibitem[{{Rein} {et~al.}(2019){Rein}, {Tamayo}, \& {Brown}}]{Rein_2019_1}
{Rein}, H., {Tamayo}, D., \& {Brown}, G. 2019, \mnras, 489, 4632

\bibitem[{{Ruth}(1983)}]{Ruth_1983}
{Ruth}, R.~D. 1983, IEEE Transactions on Nuclear Science, 30, 2669

\bibitem[{{Saillenfest}(2020)}]{Saillenfest_2020}
{Saillenfest}, M. 2020, Celestial Mechanics and Dynamical Astronomy, 132, 1

\bibitem[{{Sharina} {et~al.}(2018){Sharina}, {Ryabova}, {Maricheva}, \&
  {Gorban}}]{Sharina-Ryabova-Maricheva-Gorban_2018}
{Sharina}, M., {Ryabova}, M., {Maricheva}, M., \& {Gorban}, A. 2018, Astronomy
  Reports, 62, 733

\bibitem[{{Simon-Petit} {et~al.}(2018){Simon-Petit}, {Perez}, \&
  {Duval}}]{Simon-Petit_2018}
{Simon-Petit}, A., {Perez}, J., \& {Duval}, G.~T. 2018, Communications in
  Mathematical Physics, 363, 605

\bibitem[{{Simon-Petit} {et~al.}(2019){Simon-Petit}, {Perez}, \&
  {Plum}}]{Simon-Petit_2019}
{Simon-Petit}, A., {Perez}, J., \& {Plum}, G. 2019, MNRAS, 484, 4963

\bibitem[{{Smirnova}(2015)}]{Smirnova_2015}
{Smirnova}, L.~V. 2015, Open Astronomy, 24, 209

\bibitem[{{Wisdom} \& {Hernandez}(2015)}]{Wisdom-Hernandez_2015}
{Wisdom}, J. \& {Hernandez}, D.~M. 2015, \mnras, 453, 3015

\bibitem[{{Wisdom} \& {Holman}(1991)}]{Wisdom_1991}
{Wisdom}, J. \& {Holman}, M. 1991, \aj, 102, 1528

\bibitem[{{Yoshida}(1990)}]{Yoshida_1990}
{Yoshida}, H. 1990, Physics Letters A, 150, 262

\bibitem[{{Yoshida}(1993)}]{Yoshida_1993}
{Yoshida}, H. 1993, Celestial Mechanics and Dynamical Astronomy, 56, 27

\bibitem[{{Zotos}(2014)}]{Zotos_2014}
{Zotos}, E.~E. 2014, Mechanics Research Communications, 62, 102

\end{thebibliography}
\appendix
\section{Bounded Hénon's isochrone orbits}
\subsection{Deriving the time dependence}
\label{an:calDetailEQDIFF}
The change of variable
\begin{equation*}
    \begin{aligned}
        u=\sqrt{1+\left(r/b\right)^2}\,
    \end{aligned}
\end{equation*}
in the energy equation in Eq.~\eqref{eq:h_of_A} allows us to obtain the following non-linear differential equation:
\begin{equation}
    \label{eq:du2_dt2}
    \begin{aligned}
        \left(\mathrm{d}_t u^2\right)^2=4\left(k_3(u^2-1)+k_2(u-1)-k_1\right)\,,
    \end{aligned}
\end{equation}
where $k_1 = \left(\Lambda/b^2\right)^2$, $k_2=2\mu/b^3$, and $k_3=2h/b^2$. The next change of variable,
\begin{equation*}
    \begin{aligned}
        s=\mathbf{r}\cdot\dot{\mathbf{r}}=\sqrt{k_3(u^2-1)+k_2(u-1)-k_1}\,,
    \end{aligned}
\end{equation*}
allows us to integrate the previous equation:
\begin{equation}
    \label{eq:DtI}
    \begin{aligned}
        \Delta t = \frac{1}{k_3}\int\mathrm{d}s-\frac{k_2}{2k_3}\underbrace{\int\frac{\mathrm{d}u}{\sqrt{k_3(u^2-1)+k_2(u-1)-k_1}}}_{I}\,,
    \end{aligned}
\end{equation}
where the quantity $I$ (see the bottom curly bracket in Eq.~\eqref{eq:DtI}) can be rewritten as
\begin{equation*}
    \begin{aligned}
       I\sqrt{k_3} &= \int\frac{\mathrm{d}v}{\sqrt{v^2-k_4}} \\
       &= \ln(\sqrt{v^2-k_4}+v) \\ 
       &= \arctanh\left(\frac{v}{\sqrt{v^2-k_4}}\right) \\
       &= \arctanh\left(\frac{\mathrm{d}_u s}{\sqrt{k_3}}\right) \,.
    \end{aligned}
\end{equation*}
Here, we have defined $v=u+k_2/2k_3$ and $k_4=\left(4k_3\left(k_1+k_2+k_3\right)+k_2^2\right)/4k_3^2$. So, the time it takes for a particle to move between two positions is finally given by
\begin{equation*}
    \begin{aligned}
       \Delta t &= \frac{s}{k_3}-\frac{k_2}{2k_3^{3/2}}\arctanh\left(\frac{\mathrm{d}_u s}{\sqrt{k_3}}\right) \\
       &=\frac{\mathbf{r}\cdot\dot{\mathbf{r}}}{2h}-\frac{\mu}{\sqrt{2h^{3}}}\arctanh\left(\frac{1}{\sqrt{2h}}\mathrm{d}_{\sqrt{r^2+b^2}}\left(\mathbf{r}\cdot\dot{\mathbf{r}}\right)\right)\,.
    \end{aligned}
\end{equation*}

\subsection{Orbit classification}
\label{an:detailOrbit_h<0}
\begin{figure}
    \includegraphics[width=\columnwidth]{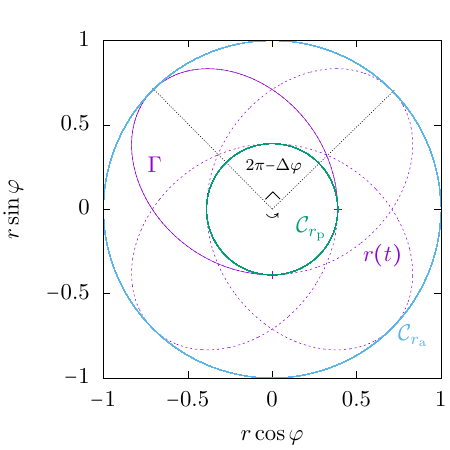}
    \caption{Example of a particle's motion in its orbital plane (a special case of a closed orbit where $\Delta\varphi=3\pi/2$) with the isochrone parameters $\mu=1$ and $b=2\times10^{-1}$. The green and blue circles, labelled $\mathcal{C}_{r_{\mathrm{p}}}$ and $\mathcal{C}_{r_{\mathrm{a}}}$, respectively show the particle's minimum and maximum radial positions. The dashed violet line, denoted by $r(t)$, shows the particle's orbit. The solid violet line shows the path of the particle during one radial period; its length is given by $\Gamma$ (see Eq.~\eqref{eq:Gamma}). The small arrow indicates the rotational sense of the loops.}
    \label{fig:RosetteForm}
\end{figure}

Figure~\ref{fig:RosetteForm} shows an example of a bounded particle’s Hénon isochrone trajectory. Bounded orbits form a series of loops, where each loop is shifted by
\begin{equation}
    \label{eq:Dphi}
    \begin{aligned}
        \Delta \varphi=\pi\left(1+\frac{\Lambda}{\sqrt{\Lambda^2+4b\mu}}\right)\,,
    \end{aligned}
\end{equation}
after each radial period. The curvilinear length of each loop is given by
\begin{equation}
    \label{eq:Gamma}
    \begin{aligned}
        \Gamma = \alpha\int_0^{2\pi}\left(1-\epsilon\cos{E}\right)\sqrt{\frac{\alpha\left(1+\epsilon\cos{E}\right)-b}{\alpha\left(1-\epsilon\cos{E}\right)+b}}~\mathrm{d}E\,.
    \end{aligned}
\end{equation}
Depending on the possible combinations of values for the particle’s orbital parameters --- semi-major axis, $\alpha$, and eccentricity, $\epsilon$ --- and the isochrone characteristic length parameter $b$~\footnote{By definition, we always have $\alpha\geq b$ for a bounded orbit.}, the geometric shape of the orbit resulting from the succession of loops differs:
\begin{itemize}
    \item When $b=0$, orbits are Keplerian ellipses.
    \item When $\alpha=b$, the eccentricity necessarily becomes zero (see Eq.~\eqref{eq:eccentricity_h<0}), implying that $\Gamma=0$: the particle remains unmoved at the origin.
    \item When $\epsilon=0$ and $0\leq b<\alpha$, Eq.~\eqref{eq:Gamma} reduces to the integral of the perimeter of a circle, and the trajectory depicts a circle with radius given in Eq.~\eqref{eq:r(E)}.
    \item When $\epsilon=1-b/\alpha$, the angular momentum is zero. Equation~\eqref{eq:Gamma} reduces to the integral of the length of a straight line, and the orbit is a line segment, as shown by Eqs.~\eqref{eq:r(E)} and~\eqref{eq:phi(E)}.
    \item The other cases are rosette-shaped, characterised by two periods: a radial and an apsidal period.
\end{itemize}

All different classes of bounded orbits are summarised in Table~\ref{tab:BoundOrbitForm}. The orbit is closed if and only if the radial and apsidal periods are multiples of each other, in which case $\Delta\varphi\in2\pi\mathbb{Q}$, which is a necessary and sufficient condition. Bertrand's theorem indicates that, in the family of all isochrone potentials (see~\citealp{Simon-Petit_2018}), only the Keplerian and harmonic potentials have all their bound orbits closed~\citep{Bertrand_1873}. Figure~\ref{fig:RosetteForm} shows a case where a bound Hénon isochrone orbit is closed.

\subsection{Polar angle}
\label{an:polaran}
For a bound orbit, the polar angle $\varphi$ as a function of the eccentric anomaly given in Eq.~\eqref{eq:phi(E)} can be rewritten as
\begin{equation*}
    \begin{aligned}
        \varphi\left(E\right)&=\frac{\Lambda}{\sqrt{\Lambda^2+4b\mu}}\arctan\left(\sqrt{\frac{\mu\alpha}{\Lambda^2+4b\mu}}\left(1+\frac{b}{\alpha}+\epsilon\right)\tan \left(\frac{E}{2}\right)\right) \\
        &\quad+\arctan\left(\frac{\sqrt{\mu\alpha}\left(1-\frac{b}{\alpha}+\epsilon\right)}{\Lambda}\tan \left(\frac{E}{2}\right)\right) \\
        &=\frac{\Lambda}{\sqrt{\Lambda^2+4b\mu}}\arctan\left(\frac{1+\frac{b}{\alpha}+\epsilon}{1+\epsilon-\frac{\sqrt{r^2+b^2}}{\alpha}}\frac{\mathbf{r}\cdot\dot{\mathbf{r}}}{\sqrt{\Lambda^2+4b\mu}}\right) \\
        &\quad+\arctan\left(\frac{1-\frac{b}{\alpha}+\epsilon}{1+\epsilon-\frac{\sqrt{r^2+b^2}}{\alpha}}\frac{\mathbf{r}\cdot\dot{\mathbf{r}}}{\Lambda}\right)\,,
    \end{aligned}
\end{equation*}
where we have used
\begin{equation*}
    \left\{
    \begin{aligned}
        \epsilon\cos{E} &= 1-\frac{\sqrt{r^2+b^2}}{\alpha} \\ \epsilon\sin{E} &= \frac{\mathbf{r}\cdot\dot{\mathbf{r}}}{\sqrt{\mu\alpha}}
    \end{aligned}
    \right.\,.
\end{equation*}
This form of the equation is more convenient when we take the limit as the energy $h$ tends towards zero in Sect.~\ref{sec:Unbound_orbit-h=0}. In this case,~$\alpha \xrightarrow[h \to 0]{} \infty$.

\section{Algorithm for the isochrone sub-steps}
\label{sec:algoiso}
In this section we outline an algorithm to propagate a particle from the state vector $(\mathbf{r}_0,\dot{\mathbf{r}}_0)$ at time $t_0$ to the state vector $(\mathbf{r},\dot{\mathbf{r}})$ at time $t=t_0+\delta t$ following the isochrone dynamics with parameters $\mu$ and $b$. We start by computing the angular momentum $\mathbf{\Lambda}$ and energy $h$ as
\begin{equation}
   \begin{aligned}
      \mathbf{\Lambda} &= \mathbf{r}_0\times\dot{\mathbf{r}}_0\,, \\
      z &= \frac{\|\dot{\mathbf{r}}_0\|}{\mu} - \frac{2}{b + \sqrt{r_0^2 + b^2}}\,.
   \end{aligned}
\end{equation}
We note that $z$ is equal to $2h/\mu$. According to the value of $z$, we select the type of trajectory followed by the particle (see Sect.~\ref{sec:IsochroneOrbit}). If $z<0$, the trajectory is bound and $z$ is equal to $-1/\alpha$. We then compute $k_0 = \epsilon\cos E_0$ and $l_0 = \epsilon\sin E_0$ as
\begin{equation}
   \begin{aligned}
      k_0 &= 1 + z\sqrt{r_0^2 + b^2}\,, \\
      l_0 &= \mathbf{r}_0\cdot\dot{\mathbf{r}}_0\sqrt{-\frac{z}{\mu}}\,,
   \end{aligned}
\end{equation}
and the variation of mean anomaly as $\delta M = \delta t\sqrt{-z^3\mu}$. We can now solve the variational Kepler equation,
\begin{equation}
   \delta M - l_0 = \delta E - k_0\sin\delta E - l_0\cos\delta E\,,
\end{equation}
for the variation $\delta E$ in eccentric anomaly. This step is realised with a root-finding algorithm that converges to machine precision within a few iterations (typically two) for any value of $k_0$, $l_0$, and $\delta M$.

At that point, as we did not find an isochrone version of the Gauss functions $f$ and $g$ (see e.g.~\citealp{Danby_1988}), a few operations still need to be performed. We first compute the unitary vector $\hat{\boldsymbol{\epsilon}}$ pointing towards the previous periapsis,
\begin{equation}
   \hat{\boldsymbol{\epsilon}} = (\hat{\mathbf{\Lambda}}\times\hat{\mathbf{r}}_0)\times\hat{\mathbf{\Lambda}}\cos\varphi_0 - \hat{\mathbf{\Lambda}}\times\hat{\mathbf{r}}_0\sin\varphi_0\,,
\end{equation}
where $\hat{\mathbf{\Lambda}} = \mathbf{\Lambda}/\Lambda$, $\hat{\mathbf{r}}_0 = \mathbf{r}_0/r_0$, and the initial true anomaly $\varphi_0$ is computed from $E_0=\mathrm{atan2}(l_0,k_0)$ using Eq.~\eqref{eq:phi(E)}. The vector $\hat{\boldsymbol{\epsilon}}$ is needed to compute the rotation matrix $\mathcal{R}$ that converts a vector expressed in the orbital reference frame (i.e. with first axis $\hat{\boldsymbol{\epsilon}}$ and third axis $\hat{\mathbf{\Lambda}}$) to the working reference frame. The three columns of matrix $\mathcal{R}$ are respectively $\hat{\boldsymbol{\epsilon}}$, $\hat{\mathbf{\Lambda}}\times\hat{\boldsymbol{\epsilon}}$, and $\hat{\mathbf{\Lambda}}$.

The new distance $r$ of the particle is computed as
\begin{equation}\label{eq:r2equiv}
   r^2 = r_0^2 + p(p - 2\sqrt{r_0^2 + b^2})\,,
\end{equation}
where
\begin{equation}
   p = \frac{1}{z}\left(2\left(1 + z\sqrt{r_0^2 + b^2}\right)\sin^2\frac{\delta E}{2} + l_0\sin\delta E\right)\,.
\end{equation}
Equation~\eqref{eq:r2equiv} is equivalent to Eq.~\eqref{eq:r(E)} but it avoids dramatic cancellation errors occurring whenever $r\ll b$. We also computed the quantity $s = \mathbf{r}\cdot\dot{\mathbf{r}}$ as
\begin{equation}
   s = \sqrt{-\frac{\mu}{z}}\epsilon\sin E\,,
\end{equation}
where $\epsilon = \sqrt{k_0^2 + l_0^2}$ and $E = E_0 + \delta E$. Expressed in the orbital reference frame, the new position vector of the particle has components $(r\cos\varphi,r\sin\varphi,0)$, and the new velocity vector has components $(s\cos\varphi - \Lambda\sin\varphi, \Lambda\cos\varphi + s\sin\varphi,0)/r$, where the true anomaly $\varphi$ is computed from $E$ using Eq.~\eqref{eq:phi(E)}. It now remains to multiply these two vectors by the matrix $\mathcal{R}$, which finally gives the position $\mathbf{r}$ and velocity $\dot{\mathbf{r}}$ of the particle in the working reference frame.

Unbound trajectories (i.e. when $h\geq0$) and special cases (e.g. when $\Lambda = 0$) are treated in a similar manner with the corresponding formulas.

\section{Properties of the function $\varepsilon\mathcal{B}_q(r)$}
\label{an:proofEpsB}
In this section we study the properties of the continuous and differentiable function $\varepsilon\mathcal{B}_q(r)$, defined in Eq.~\eqref{eq:eB0} as
\begin{equation}
    \label{eq:atler-epsB}
    \begin{aligned}
        \varepsilon\mathcal{B}_{q}(r)=\frac{q^2+2}{\sqrt{q^2+1}\left(1+\sqrt{1+r^2\left(q^2+2\right)}\right)}-\frac{1}{\sqrt{1+r^2}}\,,
    \end{aligned}
\end{equation}
where $r\geq0$ is the particle's radial position, and $q\geq0$ is the distance used as parameter.

\subsection{Positive function}
\label{an:Des_epsB>0}
We have the following demonstrated propositions:
\begin{enumerate}[1.]
    \item \begin{equation*}
          \left\{
          \begin{aligned}
                \varepsilon\mathcal{B}_{q}(r)&=0 \\
                \partial_{q}\varepsilon\mathcal{B}_{q}(r)&=0
          \end{aligned}
          \right.
          \quad\iff\quad
          \begin{aligned}
                r=q
          \end{aligned}
          \,.
          \end{equation*}
    \item As a function of $q$, the function $\varepsilon\mathcal{B}_{q}(r)$ is monotonically decreasing for $q\in(0,r)$, because $\partial_{q}\varepsilon\mathcal{B}_{q}(r)<0$ . \\
    \item As a function of $q$, the function $\varepsilon\mathcal{B}_{q}(r)$ is monotonically increasing for $q>r$, because $\partial_{q}\varepsilon\mathcal{B}_{q}(r)\geq0$.
    \item \begin{equation*}
          \begin{aligned}
                &\varepsilon\mathcal{B}_{0}(r)=\frac{2}{1+\sqrt{1+2r^2}}-\frac{1}{\sqrt{1+r^2}}\geq0\,, \\
                &\lim_{q\to\infty}\varepsilon\mathcal{B}_{q}(r)=\frac{1}{r}-\frac{1}{\sqrt{1+r^2}}>0\,.
          \end{aligned}
          \end{equation*}
\end{enumerate}
From this set of propositions, we deduce that 
\begin{equation*}
    \begin{aligned}
        \varepsilon\mathcal{B}_{q}(r)\geq 0\ \forall
(q,r)\in\mathbb{R}^2_+\,.
    \end{aligned}
\end{equation*}

\subsection{Extrema of $\varepsilon\mathcal{B}_q(r)$}
\label{an:MaxExist}
The derivative of $\varepsilon\mathcal{B}_q(r)$ with respect to $r$ is a sum of two terms (see Eq.~\eqref{eq:eB0}). By factorising the equation and multiplying the numerator by its conjugate to elevate both terms to the square, we obtain
\begin{equation}\label{eq:deBw}
    \partial_r\varepsilon\mathcal{B}_q = \frac{r(x-x_q)p_5(x)}{w}\,,
\end{equation}
where $w$ is a strictly positive quantity for any $r,q\geq 0$, and $p_5(x)$ is a polynomial of order $5$ in $x$ that does not cancel in $x=x_q$. In this expression, we have written $x=\sqrt{1+r^2(q^2+2)}$ and $x_q=1+q^2$. The expression of $p_5(x)$ is
\begin{equation}
    \begin{aligned}
        p_5(x) = &- x^5
                  + 3x_qx^4
                  + 3x_qx^3
                  + x_q(3x_q+4)x^2 \\
                 &+ x_q(x_q+1)x
                  + x_q^2(x_q+1)\,.
    \end{aligned}
\end{equation}
The function $\partial_r\varepsilon\mathcal{B}_q$ in Eq.~\eqref{eq:deBw} has the two obvious roots $r=0$ and $r=q$ (i.e. $x=x_q$). All possible remaining roots of $\varepsilon\mathcal{B}_q(r)$ are contained within $p_5(x)$. Therefore, we need to study the behaviour of $p_5(x)$ for any $x,x_q\geq 1$ (i.e. $r,q\geq 0$). First, we note that
\begin{equation}
    p_5(1) = x_q^3 + 5x_q^2 + 11x_q - 1
,\end{equation}
which is strictly positive for $x_q\geq 1$. Moreover, $p_5(x)$ tends to~$-\infty$ when $x\to +\infty$. We deduce that $p_5(x)$ has at least one root for $x\geq 1$. This means that, for any $q\geq 0$ there exists at least one additional value of $r$ (other than $r=0$ and $r=q$) for which $\partial_r\varepsilon\mathcal{B}_q$ cancels. Second, we note that the first derivative of $p_5(x)$ computed in $x=1$ is
\begin{equation}
   p_5'(1) = 7x_q^2 + 30x_q - 5\,,
\end{equation}
which is strictly positive for $x_q\geq 1$. Moreover, $p_5'(x)$ tends to~$-\infty$ when $x\to +\infty$. We deduce that $p_5(x)$ is not monotonous for $x\geq 1$, because its derivative cancels at least once. Third, the discriminant of the second derivative of $p_5(x)$ is
\begin{equation}
   \begin{aligned}
       \Delta = -3456(9x_q + 8)(6x_q + 5)^2x_q^2\,,
   \end{aligned}
\end{equation}
which is strictly negative for $x_q\geq 1$. This means that $p_5''(x)$ has only one real root. We deduce that the first derivative of $p_5(x)$ cancels exactly one time on the interval $x\geq 1$. As a result, $p_5(x)$ also has exactly one root for $x\geq 1$. Consequently, there exists exactly one additional value of $r$ (other than $r=0$ and $r=q$) for which $\partial_r\varepsilon\mathcal{B}_q$ cancels. We call this value $r_{\mathrm{c}}$, and we write $x_{\mathrm{c}} = \sqrt{1 + r_{\mathrm{c}}^2(q^2 + 2)}$. Finally, we note that
\begin{equation}
    p_5(x_q) = 2(x_q + 1)^3x_q^2\,,
\end{equation}
which is strictly positive for $x_q\geq 0$. We have therefore $x_{\mathrm{c}} > x_q$. This means that $r_{\mathrm{c}} > q$.

Incidentally, we note that $x_{\mathrm{c}}$ is a simple root of $p_5(x)$. This comes from the facts that $p_5(1)>0$, $p_5'(1)>0$, $p_5'(x)$ cancels only once, and $p_5(x)$ tends to $-\infty$ when $x\to\infty$. This property is used in Appendix~\ref{an:cancellation}.

\section{Cancellation-safe formula for implementing $\partial_r\varepsilon\mathcal{B}$}
\label{an:cancellation}
The function $\varepsilon\mathcal{B}$ in Eq.~\eqref{eq:Ham_epsB} is defined as a difference between two potentials. As we want $|\varepsilon\mathcal{B}|$ to be as small as possible, the result of this difference is generally very close to zero, which generates dramatic cancellation errors when we compute it numerically. Here, we give an equivalent expression designed to mitigate cancellation errors.

In practice, the numerical integrator only needs to compute the gradient of $\varepsilon\mathcal{B}(\mathbf{r})$ in order to update the velocity of the particle. This gradient must be computed with a cancellation-safe formula. Its expression is\begin{equation}
    \partial_\mathbf{r}\varepsilon\mathcal{B} = \partial_r\varepsilon\mathcal{B}~\partial_\mathbf{r}r\,,
\end{equation}
where
\begin{equation}
    \label{eq:deBcancel}
    \partial_r\varepsilon\mathcal{B} = r\left(\frac{\eta}{(r^2+\kappa^2)^{3/2}} - \frac{\mu}{\sqrt{r^2+b^2}(b+\sqrt{r^2+b^2})^2}\right)\,,
\end{equation}
and $\mu$ and $b$ are given in Eq.~\eqref{eq:mub}. As shown in Appendix~\ref{an:proofEpsB}, $\partial_r\varepsilon\mathcal{B}$ has exactly three roots for $r\geq 0$. The first root is $r=0$, as obtained trivially from Eq.~\eqref{eq:deBcancel}. The second root is $r=q$, by definition of $q$. The third root is written $r=r_{\mathrm{c}}$, and we know that $r_{\mathrm{c}} > q$.

The root $r=0$ does not generate cancellation errors unless $q=0$. It remains to take care of the two remaining roots, $r=q$ and $r=r_{\mathrm{c}}$. The most problematic root is $r=q$, because by definition of $q$, the location $r=q$ is necessarily reached by the particle at some point along its orbit.

We first rewrite the difference in Eq.~\eqref{eq:deBcancel} as a single fraction, and elevate both terms to the square by multiplying by their conjugate. When replacing $\mu$ and $b$ by their definitions in Eq.~\eqref{eq:mub}, we obtain
\begin{equation}\label{eq:deBfactor}
    \partial_r\varepsilon\mathcal{B} = \frac{\eta^2r(d^2 - \lambda_q^2)(\lambda-\lambda_q)P_5(\lambda)}{\lambda_q^4\lambda(d+\lambda)^2(r^2+\kappa^2)^{3/2}\left(\mu(r^2+\kappa^2)^{3/2}+\eta \lambda(d+\lambda)^2\right)} \,,
\end{equation}
where $\lambda=\sqrt{r^2+b^2}$, $\lambda_q=\sqrt{q^2+b^2}$, $d=\sqrt{q^2+\kappa^2}$, and $P_5(\lambda)$ is a polynomial of order $5$ in $\lambda$ that does not cancel in $\lambda=\lambda_q$. Cancellation errors in $r=q$ are now isolated in the term $\lambda-\lambda_q$, which we finally express as
\begin{equation}
   \lambda-\lambda_q = \frac{(r-q)(r+q)}{\lambda+\lambda_q}\,.
\end{equation}
The coefficients of the polynomial $P_5(\lambda)=a_5\lambda^5 + a_4\lambda^4 + a_3\lambda^3 + a_2\lambda^2 + a_1\lambda + a_0$ are
\begin{equation}\label{eq:Pcancel}
    \left\{
    \begin{aligned}
        a_5 &= - \lambda_q^2 \\
        a_4 &=  3\lambda_q^3 \\
        a_3 &=  3(d^2 - \lambda_q^2)\lambda_q^2 \\
        a_2 &=  (4d^2 - \lambda_q^2)(d^2 - \lambda_q^2)\lambda_q \\
        a_1 &=  d^2(d^2 - \lambda_q^2)^2 \\
        a_0 &=  d^2(d^2 - \lambda_q^2)^2\lambda_q\,.
    \end{aligned}
    \right.
\end{equation}
If $q\gg\kappa$, the term $d^2-\lambda_q^2$ can also be a source of cancellation errors. Therefore, we replace every occurrence of $d^2-\lambda_q^2$ in Eqs.~\eqref{eq:deBfactor} and~\eqref{eq:Pcancel} by
\begin{equation}
    d^2 - \lambda_q^2 = \frac{\kappa^2(\kappa^2+q^2)}{2\kappa^2+q^2}\,.
\end{equation}
It remains to take care of the cancellation errors occurring in the neighbourhood of $r=r_{\mathrm{c}}$. Unfortunately, there is no analytical expression for $r_{\mathrm{c}}$ or for the associated $\lambda_{\mathrm{c}}=\sqrt{r_{\mathrm{c}}^2+b^2}$. We can still factor the polynomial $P_5$ as $P_5(\lambda) = (\lambda-\lambda_{\mathrm{c}})P_4(\lambda)$, where $\lambda_{\mathrm{c}}$ is obtained numerically; however, numerical experiments show that this expression is extremely sensitive to the value of $\lambda$, such that it is much less numerically stable. Fortunately, cancellation errors occurring at $r\approx r_{\mathrm{c}}$ are less critical than those occurring at $r\approx q$, because not all trajectories reach this radial location. Hence, in this article we stick to Eq.~\eqref{eq:deBfactor} and leave further refinements of this expression to future works.

\begin{figure}
    \includegraphics[width=\columnwidth]{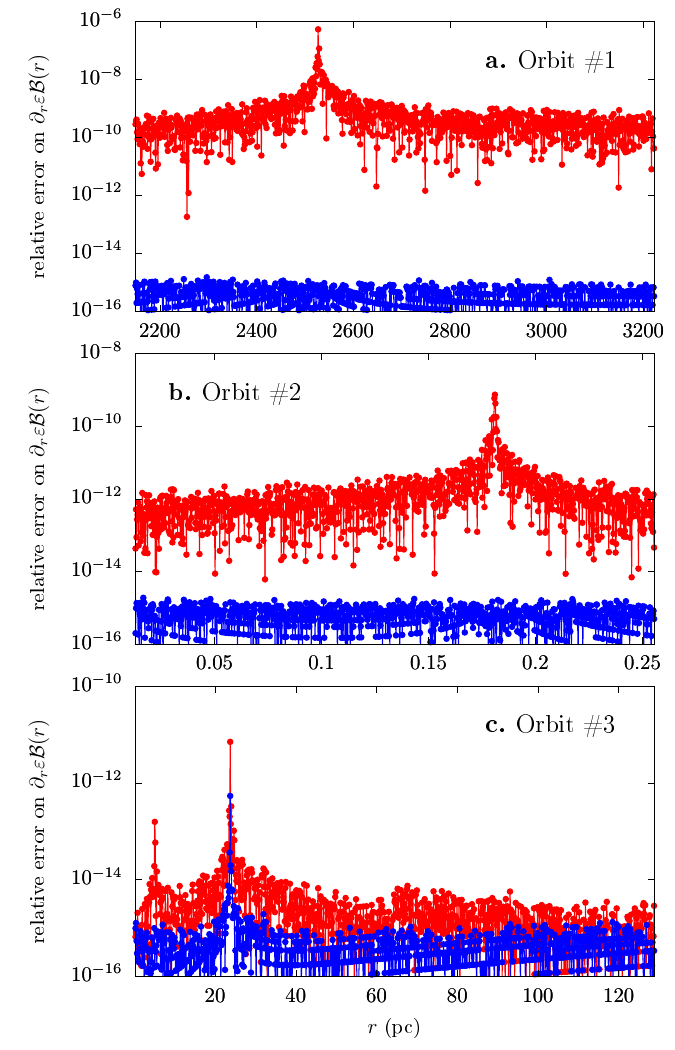}
    \caption{Numerical errors on the computation of $\partial_r\varepsilon\mathcal{B}$ using double precision arithmetic. The red curve shows the relative error obtained when using Eq.~\eqref{eq:deBcancel}; the blue curve shows the relative error obtained when using Eq.~\eqref{eq:deBfactor}. Errors are computed from a reference value obtained in quadruple precision. Panel~a,~b, and~c show the error as a function of the distance $r$ of the particle for the Orbits~$\#1$,~$\#2$, and~$\#3$ considered in Sect.~\ref{sec:perf}, with parametrisation point $q=r_\star$. The values of $r$ shown range from periapsis to apoapsis.}
    \label{fig:cancellation_rstar}
\end{figure}

Figure~\ref{fig:cancellation_rstar} compares the results obtained when using the direct difference in Eq.~\eqref{eq:deBcancel} or the factored expression in Eq.~\eqref{eq:deBfactor}. The red curve in panels~a and~b show a clear peak at $r\approx q$, where errors are maximum. These errors are absent in the blue curve, which always remain near machine precision. The red curve in panel~c features two peaks: the first peak arises near $r=q$, and it is corrected on the blue curve. The second peak arises near $r=r_{\mathrm{c}}$.

\end{document}